\documentclass[hyper,letterpaper]{JHEP3}
\usepackage{amsmath,amssymb,amsfonts,bm,amscd}
\usepackage{cite}
\usepackage{graphicx}
\usepackage{multirow}
\usepackage{verbatim}
\usepackage{appendix}
\usepackage{fancybox}
\usepackage{slashed}

\usepackage{url}
\usepackage{float}

\newcommand{\bea}{\begin{eqnarray}}
\newcommand{\eea}{\end{eqnarray}}
\newcommand{\be}{\begin{equation}}
\newcommand{\ee}{\end{equation}}

\newcommand{\Z}{{\mathbb Z}}
\newcommand{\R}{{\mathbb R}}
\newcommand{\C}{{\mathbb C}}

\def\Tr{{\rm Tr \,}}

\def\frak{\mathfrak}

\def\tilde{\widetilde}
\def\hat{\widehat}
\def\bar{\overline}

\def\CC{{\mathcal C}}

\def\CH{{\mathcal H}}
\def\CI{{\mathcal I}}

\def\CL{{\mathcal L}}
\def\CM{{\mathcal M}}
\def\CN{{\mathcal N}}
\def\CO{{\mathcal O}}

\def\CR{{\mathcal R}}
\def\CS{{\mathcal S}}
\def\CT{{\mathcal T}}

\newcommand{\cp}{{\mathbb{C}}{\mathbf{P}}}

\renewcommand{\bar}{\overline}
\renewcommand{\hat}{\widehat}

\title{Counting RG flows}

\author{Sergei Gukov$^{1,2}$
\\
$^1$ Walter Burke Institute for Theoretical Physics, California Institute of Technology, Pasadena, CA 91125 USA \\
$^2$ Max-Planck-Institut f\"ur Mathematik, Vivatsgasse 7, D-53111 Bonn, Germany}

\abstract{Interpreting renormalization group flows as solitons interpolating between different fixed points,
we ask various questions that are normally asked in soliton physics but not in renormalization theory.
Can one count RG flows?
Are there different ``topological sectors'' for RG flows?
What is the moduli space of an RG flow,
and how does it compare to familiar moduli spaces of (supersymmetric) dowain walls?
Analyzing these questions in a wide variety of contexts --- from counting RG walls to AdS/CFT correspondence ---
will not only provide favorable answers, but will also lead us to a unified general framework
that is powerful enough to account for peculiar RG flows and predict new physical phenomena.
Namely, using Bott's version of Morse theory we relate the topology of conformal manifolds to certain properties
of RG flows that can be used as precise diagnostics and ``topological obstructions'' for the strong form of the $C$-theorem in any dimension.
Moreover, this framework suggests a precise mechanism for how the violation of the strong $C$-theorem happens
and predicts ``phase transitions'' along the RG flow when the topological obstruction is non-trivial.
Along the way, we also find new conformal manifolds in well-known 4d CFT's
and point out connections with the superconformal index and classifying spaces of global symmetry groups.
\\
\\
\\
\\
\\
\\
\\
{\tt CALT-TH-2015-010}}

\begin{document}
\cornersize{1}


\section{Introduction}
\label{sec:intro}

The renormalization group (RG) flow is a one-parameter motion \cite{Wilson:1973jj} in the space of (renormalized) coupling constants $\lambda^i$,
\be
\frac{d}{dt} \equiv - \beta^i (\lambda) \frac{\partial}{\partial \lambda^i}
\ee
with beta-functions $\beta^i (\lambda)$ as ``velocities'' and the RG ``time'' $t= - \log \mu$ increasing toward the infra-red (IR).
The fixed points of the flow are conformal field theories (CFTs) that we denote by $T_{\text{UV}}$ (when the flow originates at $T_{\text{UV}}$)
or by $T_{\text{IR}}$ (when the flow ends at $T_{\text{IR}}$) or simply by $T_*$ when we wish to make a statement about a general fixed point of the RG flow.

\begin{figure}[ht] \centering
\includegraphics[width=2.3in]{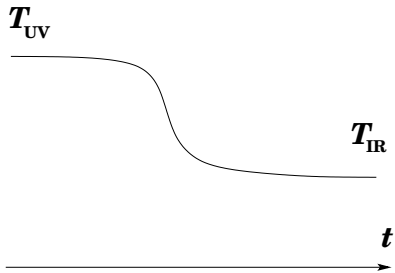}
\caption{\label{fig:RGflows2}A schematic representation of a renormalization group (RG) flow from a UV conformal theory $T_{\text{UV}}$ to an IR conformal theory $T_{\text{IR}}$. The flow is parametrized by $t \in (- \infty, + \infty)$.}
\end{figure}

Interested in RG flows from a CFT $T_{\text{UV}}$ to another CFT $T_{\text{IR}}$, we denote by $\CT$ the space of theories in which the RG trajectories $\lambda^i (t)$ are embedded. For instance, $\CT$ can be the space of all $d$-dimensional Quantum Field Theories (if $T_{\text{UV}}$ and $T_{\text{IR}}$ are both $d$-dimensional CFTs).\footnote{In some cases, RG flows go across dimensions. Our analysis still applies to such situations, where $T_{\text{UV}}$ and $T_{\text{IR}}$ are theories in different dimensions. One simply needs to pick the smallest dimension and regard all other theories --- including the intermediate stages of the RG flow --- as theories in that smallest dimension (possibly, with infinitely many fields).} Or, more economically, if RG flows preserve certain symmetry and / or supersymmetry of the fixed points $T_{\text{UV}}$ and $T_{\text{IR}}$, then we can choose $\CT$ to be the space of all theories with such properties, {\it e.g.} the space of 4d $\CN=1$ theories. Clearly, all such choices provide us with infinite-dimensional theory space $\CT$.
We can choose to truncate $\CT$ even further, to a finite-dimensional class of theories, say, parametrized by a set of couplings $\lambda^i$, $i = 1, \ldots, \dim (\CT)$. In all cases, whether $\CT$ is finite-dimensional or not, we treat $\lambda^i$ as coordinates on $\CT$.

Then, we can equivalently formulate the renormalization group flow
as a one-parameter flow on the theory space $\CT$:
\bea
\R & \to & \CT \\
t & \mapsto & T_t \nonumber
\eea
generated by a vector field $\beta$.
It has important physical property that a theory $T_t \in \CT$ with space-time metric $e^{-t} g_{\mu \nu}$
predicts the same measurements as $T_0$ with the metric $g_{\mu \nu}$.

Since renormalization group equations are PDE's in the space of couplings,
we can count their solutions much like solutions to any other system of PDE's.
What we really want, however, is a simple and effective tool
that allows to count RG flows between CFTs $T_{\text{UV}}$ and $T_{\text{IR}}$
without doing all the hard work of solving the PDEs directly.
As in other familiar enumerative problems, our goal will be to infer some information
about such RG flows from certain data of the endpoints $T_{\text{UV}}$ and $T_{\text{IR}}$
and, furthermore, relate it to the topology of the space $\CT$.

In many situations, RG flow is a gradient flow. For example, the RG flow in four-dimensional $\phi^4$ theory
is a gradient flow up to three loop order \cite{Wallace:1974dy}.
One of the consequences of the gradient flow is the existence of a ``height'' function $\CC : \CT \to \R$
that is monotonically decreasing along the flow,
\be
\frac{d \CC (T)}{dt} \le 0
\label{decreasingC}
\ee
and is stationary at the fixed points:
\be
\beta (\lambda_*) = 0 \qquad \leftrightarrow \qquad \frac{d \CC (T)}{dt} = 0
\label{stationaryC}
\ee
One can further adjust the definition of $\CC (T)$ so that at fixed points it equals the central charge of the corresponding CFT, $\CC (T_*) = c (T_*)$, the quantity that counts the number of massless degrees of freedom in a CFT $T_*$.
There are many situations where a candidate for the $\CC$-functional is known.
Such favorable situations include the celebrated Zamolodchikov's $c$-theorem \cite{Zamolodchikov:1986gt}
(in which case $\CT$ is the space of 2d QFTs) and the ``$a$-theorem'' conjecture \cite{Cardy:1988cwa} in four dimensions,
where $a$ is the coefficient of the Euler density that appears in the conformal anomaly.

To be more precise, one usually distinguishes between three versions of the ``$\CC$-theorem'' conjecture
(nicely summarized {\it e.g.} in \cite{Barnes:2004jj}).
The weakest version only involves the values of $\CC$ at the endpoints of the RG flow and asserts that $\CC (T_{\text{UV}}) > \CC (T_{\text{IR}})$.
A stronger version requires $\CC (T)$ to exist throughout the entire RG flow and states that
it is a monotonically decreasing function of $t$, that obeys \eqref{decreasingC}-\eqref{stationaryC}.
The strongest form of the ``$\CC$-theorem'' asserts that RG flow is a gradient flow of the $\CC$-function.
The first two versions are now proven theorems in two dimensions \cite{Zamolodchikov:1986gt}
and in four dimensions \cite{Komargodski:2011vj},
while the third, strongest form remains challenging even in two dimensions \cite{Ryu:2007zz},
despite the proof in conformal perturbation theory \cite{Zamolodchikov:1986gt}
and the compelling non-perturbative arguments \cite{Friedan:2009ik} that apply to a large class of RG flows.

Similarly, in four dimensions there is substantial evidence for the strongest form of the $\CC$-theorem
in a variety of theories --- {\it e.g.} in the presence of $\CN=1$ supersymmetry \cite{Barnes:2004jj,Green:2010da} or in conformal theories
that admit a holographic dual \cite{Anselmi:2000fu} --- and there are no razor-sharp counterexamples or arguments to the contrary.
In fact, {\it a priori} it is not even clear what such ``an argument to the contrary'' might look like.
It is easy to imagine how certain RG flows can rule out particular candidates for the $\CC$-function, such as the $a$-function.
But how does one show that no $\CC$-function with the desired properties can exist? ... unless there is something special
about gradient RG flows and they imply certain relations
on the data at the endpoints which, thereby, can be used as diagnostics for the behavior {\it along} the flow.
As we explain below, this is indeed the case and such local data at the endpoints $T_{\text{UV}}$ and $T_{\text{IR}}$
involves a careful count of relevant and marginal operators, together with global symmetries;
moreover, the same data is needed to answer the above questions about ``counting'' RG flows (without doing the actual count).

Precisely this problem one encounters in Morse theory or, to be more precise,
in a variant of Morse theory developed by Bott \cite{MR0064399} that will be needed for proper understanding of RG flows.
Indeed, the main goal of Morse-Bott theory is to learn about topology of the ambient space $\CT$ from
certain data at the critical points (or, critical manifolds) of the Morse function $\CC : \CT \to \R$,
without using any detailed knowledge about $\CC (T)$, especially away from the critical points.
Our goal is slightly different --- namely, to learn about the behavior of flows themselves --- but we can use the same set of tools.
Note, promoting coupling constants $\lambda_i$ to background fields in the customary fashion \cite{Seiberg:1993vc}
would result in a more familiar version of Morse theory \cite{Witten:1982im} in the space $\CT$ of (background) fields.

Taking this motivation more seriously,
and building on earlier applications of Morse theory to 2d theories \cite{MR702806,Das:1988vd,Vafa:1988ue}
and to holographic RG flows \cite{Behrndt:2001qa},
we develop a general framework that is detailed enough to teach us useful lessons about the physics of RG flows:
make predictions about new conformal manifolds, account for dangerously irrelevant operators, and not only
provide diagnostics for violation of the strong $\CC$-theorem, but also give us a hint on {\it how} and {\it why} such violations happen.
We start by testing the following three ``conjectures'' about a simple and easily accessible quantity
\be
\mu (T_*) = \# (\text{relevant}~\CO)
\label{mudef}
\ee
that counts relevant operators at the fixed points of RG flows.
The three ``conjectures'' are somewhat analogous to the three versions of the $\CC$-theorem:

\begin{enumerate}

\item
The weakest version asserts that the positive integer \eqref{mudef} counts degrees of freedom at the fixed points
and, much like the $\CC$-function, should decrease along the RG flow:
\be
\mu (T_{\text{UV}}) \; > \; \mu (T_{\text{IR}})
\label{muthma}
\ee

\item
A slightly stronger version is that, if $T_{\text{UV}}$ and $T_{\text{IR}}$ are connected by an RG flow,
then the (moduli) space of such flows has dimension:
\be
\dim \CM (T_{\text{UV}},T_{\text{IR}}) \; = \; \mu (T_{\text{UV}}) - \mu (T_{\text{IR}})
\label{muthmb}
\ee
In particular, it incorporates the weaker version as the statement that $\CM = \emptyset$ when $\mu_{\text{IR}} > \mu_{\text{UV}}$
and gives a partial answer to the question about counting RG flows.

\item
The strongest version can be formulated in situations where one has a $\CC$-function(al)
on the theory (mini-super)space $\CT$.
In such cases, the strongest version of \eqref{muthma} and \eqref{muthmb} is the conjecture:
\be
\mu (T_*) \; = \; \# \big( \text{negative eigenvalues of Hess} (\CC)\vert_{T=T_*} \big)
\label{muthmc}
\ee
where
$\text{Hess} (\CC) = \left( \frac{\partial^2 \CC}{\partial \lambda_i \partial \lambda_j} \right)$
is the Hessian of the $\CC$-function.

\end{enumerate}

\noindent
The way they are formulated, ``conjectures'' \eqref{muthma}-\eqref{muthmc} are a bit too naive and, at best,
can only apply to {\it non-degenerate} critical points, {\it i.e.} isolated (rigid) CFTs with no marginal operators or symmetries.
Therefore, in the next section, we start by introducing proper machinery based on Bott's version of Morse theory
that, among other things, will allow us to analyze CFTs with marginal deformations and non-trivial conformal manifolds.
In particular, it will provide answers to the questions posed in \cite{Kol:2002zt,Kol:2010ub} about topology of conformal manifolds.
(Free theories and theories with moduli spaces of vacua are perhaps most subtle from this perspective;
we shall try to avoid them in the present paper and explore more fully in the future work.)

\begin{figure}[ht] \centering
\includegraphics[width=4.7in]{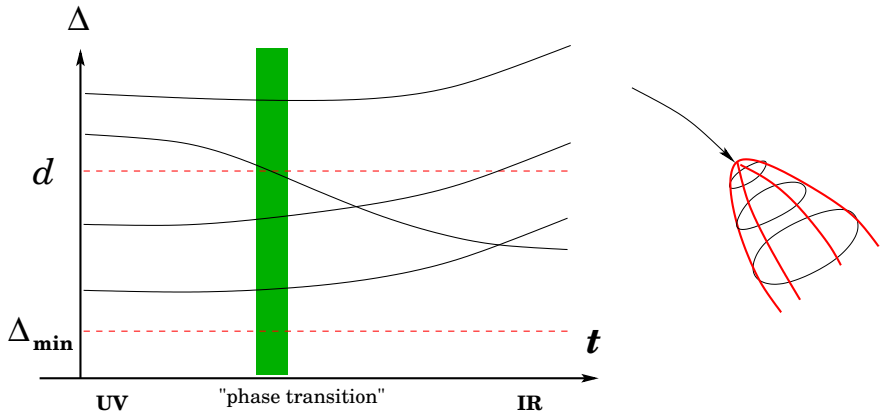}
\caption{\label{fig:RGflows4}The ``spectral flow'' of scaling dimensions in CFT$_d$ and
a cartoon illustrating how the space of flows fails to be a manifold at the ``phase transition'' point.}
\end{figure}

Another, more serious reason to put quotation marks around ``conjectures'' \eqref{muthma}-\eqref{muthmc}
and their appropriate refinements in section \ref{sec:topology} is that,
while counterexamples are rare, they do exist (and are extremely interesting!).
In particular, dangerous irrelevant operators \cite{Amit:1982az}
that play an important role in $\CN=1$ dualities \cite{Kutasov:1995ve,Kutasov:1995np,Intriligator:1995ff,Kutasov:1995ss,Leigh:1996ds}
start their life as irrelevant operators in $T_{\text{UV}}$ and, upon the RG flow, become relevant in $T_{\text{IR}}$.
When there are sufficiently many such operators, they can lead to $\mu_{\text{IR}} > \mu_{\text{UV}}$
even when the space $\CM (T_{\text{IR}},T_{\text{UV}})$ is non-empty.
Luckily, a closer look at such flows through the looking glass of Morse-Bott theory suggests what happens:
such behavior can not occur for a gradient flow indicating a ``phase transition'' at points
where anomalous dimensions of irrelevant operators cross through marginality.
At such points, new directions for the RG flow open up, which is precisely where $\CM (T_{\text{IR}},T_{\text{UV}})$ fails to be a manifold,
{\it cf.} Figure~\ref{fig:RGflows4}.
As a result, certain derivatives of the $\CC$-function (and, perhaps, some other quantities) are discontinuous at such points.
Therefore, what we really wish to conjecture is the following:

\medskip
\noindent
{\bf $\mu$-Theorem:} {\it A gradient RG flow ``breaks'' at the specific points along the flow,
which are precisely the points where irrelevant operators cross through marginality.
In the absence of such ``phase transitions'' the topological relations \eqref{muthma}-\eqref{muthmc}
(or, their appropriate variants in section \ref{sec:topology}) must hold.}

\medskip

Therefore, we can use appropriate refinements of \eqref{muthma}-\eqref{muthmc} presented in section \ref{sec:topology}
as diagnostics for the strong version of the $\CC$-theorem and whether one should expect such ``phase transitions'' or not.
We collectively refer to such topological relations as the ``$\mu$-theorem'',
even though in practice it often involves not only the index \eqref{mudef} but also
topology of conformal manifolds at the fixed points and other data.
Notice, our proposal for the piecewise structure of the gradient flow fits well with the known properties
of 2d RG flows \cite{Zamolodchikov:1986gt,Friedan:2009ik} and 4d $\CN=1$ flows \cite{Green:2010da,Amariti:2012wc}.
In particular, the ``branch flip'' in $a$-maximization is precisely an example of such ``phase transition''
and non-smooth behavior of the $a$-function.

The proposed piecewise structure of gradient flows is also reminiscent of ``wall crossing'' phenomena,
where instead of the spectrum of relevant operators one talks about spectrum of BPS states,
and the role of anomalous dimensions is played by masses of BPS states.
The latter change continuously, as piecewise smooth functions, until parameters of the theory hit a ``wall''
and the spectrum of states changes. It would be interesting to come up with a model that makes this analogy precise.\\

There are several familiar instances where counting RG flows turns into counting of BPS solitons on the nose.
These include supersymmetric domain walls that implement RG flows
as well as gravitational BPS solitons in the holographic dual description.
Although these systems are quite different, the lesson is universal,
and we will try to illustrate each of these with the simplest possible examples.
Thus, in section \ref{sec:RGwalls} we discuss counting RG walls in the setup of the pioneering work \cite{Brunner:2007ur}
where such objects were first studied.
And, in section \ref{sec:2d}, we discuss holographic RG flows in what
one might call\footnote{following how Nima Arkani-Hamed likes to call $\CN=4$ super-Yang-Mills}
the ``simplest gauged supergravity''.
In part, our motivation for choosing this holographic model is to understand RG flows
in recently discovered 2d $\CN=(0,2)$ theories~\cite{Gadde:2014ppa,Bertolini:2014ela}.
The main motivation, though, to focus on this gauged supergravity is that, despite being the simplest, it is not well studied.
We wish to emphasize that it has a rich and simple structure of phases, analogous to that of $\CN=2$ theories in two dimensions \cite{Witten:1993yc}.

Then, in sections \ref{sec:3d} and \ref{sec:4d} we illustrate the counting of RG flows in 3d and 4d theories, respectively.
As our examples we choose well known CFTs, many of which have marginal deformations and non-trivial conformal manifolds.
In some examples, we find new conformal manifolds and point out how ``topological obstructions'' discussed in this paper can be
reformulated in terms of the superconformal index.
In the case of 4d $\CN=1$ flows, we explain how ``branch flips'' in $a$-maximization fit in this general framework
and can be predicted based on topological criteria and the data at the endpoints of the RG flow.
In those example, the ``phase transition'' is of second order.


\section{Reconstructing theory space from RG flows}
\label{sec:topology}

In this section we discuss deformation-equivalence of RG flows.
In particular, our motivation comes from such questions as ``Are there distinct topological sectors for RG flows?''
and, if so, ``What invariants can distinguish different flows?''
Since RG flow is a continuous flow in the theory space $\CT$, of course, $\pi_0 (\CT)$ gives rise to different homotopy types of RG flows.
Moreover, since RG flows are one-dimensional trajectories in $\CT$, the fundamental group $\pi_1 (\CT)$ classifies different homotopy types of such paths.

\begin{figure}[ht] \centering
\includegraphics[width=2.3in]{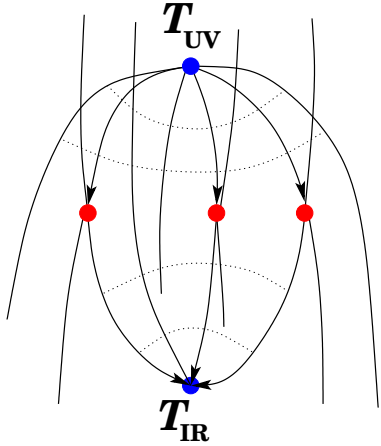}
\caption{\label{fig:RGflows1}Stable and unstable manifolds, $\CI(T_*)$ and $\CR(T_*)$,
are spanned by RG flow trajectories in and out of the CFT $T_*$. The intersection of $\CR(T_{\text{UV}})$ and $\CI(T_{\text{IR}})$ defines the set of RG flows from $T_{\text{UV}}$ to $T_{\text{IR}}$ in the theory space $\CT$.}
\end{figure}

To learn more about the relation between RG flows and topology of $\CT$
we define the {\it irrelevant manifold} of a CFT $T_*$ to be the set of theories
which flow `down' to $T_*$, {\it i.e.} for which $T_*$ is the IR fixed point:
\be
\CI (T_*) \; := \; \{ T_t \in \CT ~\vert~ \lim_{t \to + \infty} T_t = T_* \}
\label{stabmfld}
\ee
Here, for simplicity, we assume that $T_*$ is an isolated fixed point, {\it i.e.} has no exactly marginal deformations;
the relaxing of this assumption will be discussed later in this section.
Similarly, the {\it relevant manifold} $\CR (T_*)$ is the set of theories which flow `up' to $T_*$, {\it i.e.}
\be
\CR (T_*) \; := \; \{ T_t \in \CT ~\vert~ \lim_{t \to - \infty} T_t = T_* \}
\label{unstab}
\ee
With this definition of spaces $\CI (T_*)$ and $\CR (T_*)$, illustrated in Figure~\ref{fig:RGflows1},
we can be a little more precise about what we mean by the moduli space of RG flows
between $T_{\text{UV}}$ and $T_{\text{IR}}$ in \eqref{muthmb} and define it as
\be
\CM (T_{\text{UV}},T_{\text{IR}}) \; =  \; \CR (T_{\text{UV}}) \cap \CI (T_{\text{IR}})
\label{Mflows}
\ee
Even though in the context of quantum field theory $\CI(T_*)$ and $\CR(T_*)$ may be infinite-dimensional,
the space of flows \eqref{Mflows} is typically finite-dimensional.
When $\CR(T_{\text{UV}})$ and $\CI(T_{\text{IR}})$ intersect transversely, we have
\be
\dim \CM (T_{\text{IR}},T_{\text{UV}}) = \mu (T_{\text{UV}}) - \mu (T_{\text{IR}})
\label{dimMTT}
\ee
where $\mu (T_*)$ is defined\footnote{Equivalently, one could try to define $\mu$ as minus the number of irrelevant operators,
{\it i.e.} as the codimension of $\CI(T_*)$.
Clearly, it is more convenient to count relevant operators than irrelevant ones.}
as the dimension of $\CR(T_*)$:
\be
\mu (T_*) = \# (\text{relevant}~\CO)
\label{mudefb}
\ee
Note, the difference in \eqref{dimMTT} can be interpreted as the spectral flow
of the dilatation operator $\frak{D}$ that measures scaling dimensions $\Delta (\CO)$ of local operators:
\be
\frak{D} \CO \; = \; \Delta \CO
\label{dilopDO}
\ee
Since in $d$ space-time dimensions \eqref{mudefb} counts the number of operators with $\Delta (\CO) < d$,
the difference in \eqref{dimMTT} becomes the ``spectral flow'' of the operator $\frak{D} - d \cdot {\bf 1}$,
{\it i.e.} the number of eigenvalues that move from negative to positive minus the number that move from positive to negative.
An index formula for a family of elliptic operators labeled by $t \in (- \infty, + \infty)$
with a real, discrete spectrum usually relates the ``spectral side'' and the ``geometric side.''
Therefore, one can interpret \eqref{dimMTT} as an index formula for the dilation operator,
with the geometric left-hand side and the spectral right-hand side.

Let us illustrate these definitions with a concrete example.
A simple family of isolated SCFT's labelled by an integer $n=2, 3, \ldots$ is the set of $\CN=2$ super-Virasoro minimal models in two dimensions.
The $A$-type minimal model at level $n-2$ has central charge $c = 3 - \frac{6}{n}$ and $n-2$ supersymmetric relevant operators, which in the Landau-Ginzburg description \cite{Zamolodchikov:1986db,Kastor:1988ef,Martinec:1988zu,Vafa:1988uu} of the minimal model correspond to the deformation of the superpotential:
\be
W = \Phi^{n} \quad \leadsto \quad W + \delta W = \Phi^{n} + \lambda \Phi^{k} \,, \quad k<n
\label{SUSYminLG}
\ee
Each of these relevant deformations induces a supersymmetric RG flow to a $\CN=2$ minimal model at level $k-2$. Hence, in this example, one can take $\CT$ to be the space of two-dimensional $\CN=2$ theories and $\mu (A_{m}) = 2m$ for the $A_{m}$ minimal model. In particular, the coefficients of relevant perturbations are complex-valued and the space of supersymmetric RG flows has real dimension $2(n-k)$, in accord with \eqref{dimMTT},
\be
\CM (T_{\text{IR}},T_{\text{UV}}) = \C^* \times \C^{n-k-1}
\label{MflowsSUSYMM}
\ee
parametrized by complex coefficients of the relevant perturbations $\CO_m = \Phi^{m}$, $m=k, \ldots, n-1$, the first of which must be non-zero (to ensure the flow to the $A_{k-2}$ minimal model).
Note, the theory of a free (super)field $\Phi$ can flow to any of the $A_{n-2}$ minimal models and has $\mu = \infty$.
This illustrates why free theories are not so simple in this approach.
Also note that the theory with $n=2$ is trivial in the IR (has a unique massive vacuum).

More generally, in any dimension we allow $T_{\text{IR}}$ to be a vacuum sector of a massive theory ({\it e.g.} confining theory) described by a TQFT (with $\CC = 0$). According to \eqref{mudefb}, for RG flows that end at such ``trivial'' points we clearly have
\be
\mu (T_{\text{IR}})=0 \,.
\ee

The factor of $\C^* \cong S^1 \times \R$ in \eqref{MflowsSUSYMM} illustrates another general fact.
Namely, a spontaneous breaking of the scaling invariance in the RG flow
results in a collective coordinate parametrized by $t \in \R$. For this reason, a factor of $\R$ is always a part of $\CM (T_{\text{IR}},T_{\text{UV}})$ and motivates the definition of the ``reduced moduli space'' $\tilde \CM (T_{\text{IR}},T_{\text{UV}})$, such that
\be
\CM (T_{\text{IR}},T_{\text{UV}}) = \R \times \tilde \CM (T_{\text{IR}},T_{\text{UV}})
\ee

When the RG flow in question is a gradient flow for some function(al) $\CC (T)$,
then $\CI(T_*)$ and $\CR(T_*)$ are respectively stable and unstable manifolds of a critical point $T_*$.
Moreover, $\mu (T_*)$ is then the familiar definition of the Morse index and \eqref{dimMTT} becomes
the familiar formula in Morse theory.
In fact, many results and conjectures in this paper are almost automatic for RG flows which are gradient flows.
The point, though, is that all of the definitions \eqref{stabmfld}-\eqref{mudefb} do not explicitly
refer to the $\CC$-functional and, therefore, can be used in all situations, even when RG flows are not gradient flows.
{}From this perspective, counterexamples to the ``conjectures'' \eqref{muthma}-\eqref{muthmc}
and the statements below are obstructions to the existence of a $\CC$-function for which the RG flow is a gradient flow.

If we were in the context of Morse theory, {\it i.e.} if RG flows were indeed gradient flows for some $C^{\infty}$ function $\CC : \CT \to \R$,
that for now we assume to have only non-degenerate critical points ({\it cf.} \cite{Das:1988vd,Vafa:1988ue}),
then we could infer a lot more information about the ``geography'' of the theory space from the critical points of $\CC$.
In particular, it follows from the Morse lemma that relevant manifolds $\CR (T_*)$
are homeomorphic to $\R^{\mu}$ 
and that $\CC$-function provides a decomposition of $\CT$ into `cells':
\be
\CT = \bigcup_{T_*~\text{critical}} \CI (T_*) = \bigcup_{T_*~\text{critical}} \CR (T_*)
\label{TviaIandR}
\ee
By intersecting these two decompositions, we obtain a finer one,
\be
\CT = \bigcup_{T_1,T_2~\text{critical}} \CM (T_1,T_2) \,,
\label{TviaflowM}
\ee
with the moduli spaces $\CM (T_1,T_2)$ defined as in \eqref{Mflows}.
The main theorem of Morse theory gives information about how these pieces fit together:
$\CT$ has the homotopy type of a cell complex, with one cell of dimension $\mu$ for each critical point of index $\mu$.
Of course, reconstructing the space of {\it all} theories is too ambitious, and this is not what we are interested in.
For us, $\CT$ will be something extremely concrete: it will be a small ``patch'' of the space of theories that includes
all theories visited by RG flows that we wish to study.
Hence, we regard \eqref{TviaflowM} merely as a convenient geometric way to package information about RG flows.

Conversely, it means that if $\CT$ has non-trivial topology, then there must exist non-trivial CFTs in $\CT$
(or, else, we get a counterexample to the strongest version of the $\CC$-theorem).
Moreover, one can make specific predictions about CFTs just from basic information about topology of $\CT$.
Thus, the $k$-th Betti number of $\CT$ gives a lower bound on the number of CFTs in $\CT$ with $\mu=k$,
\be
N_k \ge b_k (\CT) \,,
\label{weakMorse}
\ee
where $b_k = \dim H^k (\CT ; \R)$ and
\be
N_k \; := \; \# \left( \text{critical points of index}~\mu=k \right) \,.
\ee
The bound \eqref{weakMorse} is what in Morse theory goes by the name of weak Morse inequalities.
The strong Morse inequalities in our context would look like
\be
\sum_{k=0}^n (-1)^{n-k} N_k  \; \ge \; \sum_{k=0}^n (-1)^{n-k} b_k (\CT)
\label{strongMorse}
\ee
Sometimes one writes Morse inequalities as
\be
\sum_{k} N_k q^k \; = \; \sum_{k} b_k q^k + (1 + q) \sum_{k} q^k Q_k
\label{Morseineq}
\ee
where $Q_k$ are non-negative integers (that depend on $\CC$).
Of course, the first sum on the right-hand side $P(q) = \sum_k b_k q^k$ is nothing but the Poincar\'e polynomial of $\CT$.
It can be convenient to combine the second sum also into a polynomial (or power series)
with non-negative coefficients $Q(q) = \sum_{k} q^k Q_k$.
Setting $q=-1$ in \eqref{Morseineq} gives the Morse theorem:
\be
\chi (\CT) \; = \; \sum_k (-1)^k N_k \,.
\label{MorseEuler}
\ee

The beauty of the relations \eqref{TviaIandR}-\eqref{MorseEuler} between RG flows and topology of $\CT$ is that,
while they ought to hold true if the strongest version of the $\CC$-theorem holds,
they do not explicitly refer to the $\CC$-function.
Hence, one can explore \eqref{TviaIandR}-\eqref{MorseEuler} regardless of whether the $\CC$-function exists (and is known) or not.
This will be the philosophy of the present paper.


\subsection{Conformal manifolds and RG flows}
\label{sec:Bott}

So far, our inspiration from Morse theory was based on the case where each critical point of the function(al) $\CC$ is isolated.
While it may be the case in simple examples, this is not the generic situation, of course.
In particular, we will need a variant of Morse theory developed by Bott \cite{MR0064399}
that deals with situations when this is not the case,
{\it i.e.} when $\text{Hess} (\CC)$ has zero eigenvalues and there are critical submanifolds (rather than critical points) of $\CC$.

Such situations have nice physical counterparts in our story. Namely, vanishing eigenvalues of $\text{Hess} (\CC)$
at a critical point $T_* \subset \CT$ indicate that a CFT $T_*$ has marginal operators
(that correspond to the 0-eigenspace of $\text{Hess} (\CC)$)
and critical manifolds of $\CC$ are the so-called {\it conformal manifolds}.
We call the space of conformal field theories $\CS_{\text{CFT}}$ (no pun intended) or simply $\CS (T_*)$ when
we wish to focus on conformal theories connected to $T_*$.
By definition, $\CS (T_*)$ is a submanifold of $\CT$ and much of the previous discussion
has a suitable generalization in which non-degenerate critical points $T_*$ are replaced by conformal submanifolds $\CS_* \subset \CT$.

In particular, for any $T_* \subset \CS_{\text{CFT}}$ we can continue using the definition \eqref{mudef}
of the index $\mu$ that counts relevant operators in the CFT $T_*$.
When $\CT$ is equipped with a smooth $\CC$-function, we can choose local coordinates in the space of couplings,
such that $T_*$ corresponds to $\lambda_i = 0$ and in the vicinity of this point
\be
\CC (\lambda) = \CC (T_*) + \sum_i a_i (\lambda_i)^2 + \ldots \,,
\label{CnearT}
\ee
where each $a_i$ is $-1$, $0$, or $1$.
The number of negative $a_i$'s is precisely the index $\mu (T_*)$,
which now allows to apply \eqref{muthmc} to conformal manifolds.
In the context of Morse theory, such $\CC : \CT \to \R$ is called Morse-Bott function and it reduces
to the earlier definition of the Morse function when none of the $a_i$'s are zero at every critical point of $\CC$.
In general, the set of CFTs in $\CT$ is a disjoint union of submanifolds of various (co-)dimension,
which are precisely the critical submanifolds of $\CC$.
In the above coordinate chart, the conformal submanifold $\CS_{\text{CFT}}$
is given by the equations $\lambda_i = 0$ for all $i$ with $a_i \ne 0$.

One can also relate the geometry of conformal manifolds to the geometry of RG flows, generalizing \eqref{TviaIandR}-\eqref{MorseEuler}.
In particular, now $\CT$ has the homotopy type of a {\it cell-bundle complex}, so that a generalization of \eqref{TviaIandR} looks like
\be
\CT \; = \; \bigcup_{\CS_i~\text{conformal}} \CR (\CS_i) \,,
\label{TviaRS}
\ee
where each component $\CS_i$ of the conformal manifold with index $\mu_i := \mu (\CS_i)$
contributes a cell-bundle $\CR (\CS_i)$ of rank $\mu_i$, {\it i.e.} a fiber bundle over $\CS_i$ with fiber $\R^{\mu_i}$, {\it cf.} \eqref{unstab}.
Similarly, the suitable generalization of \eqref{Morseineq} has the form
\be
\sum_{i,j} q^{\mu_i + j} \dim H^j (\CS_i ; \R) \; = \; (1+q) Q(q) + \sum_k q^k \dim H^k (\CT ; \R)
\label{MorseBottineq}
\ee
where $Q(q)$ is a polynomial with non-negative coefficients, {\it cf.} the discussion that follows~\eqref{Morseineq}.
One can write this relation more succinctly as
\be
MB(q) - P(q) = (1+q) Q(q)
\label{MorseBottineqx}
\ee
where $P(q)$ is the Poincar\'e polynomial of $\CT$
and the {\it Morse-Bott polynomial} $MB(q)$ is the sum on the left-hand side of \eqref{MorseBottineq}.
In Morse theory, the philosophy behind \eqref{MorseBottineq} is that under a slight deformation
of the Morse-Bott function $\CC$ (that turns it into a Morse function),
each critical submanifold $\CS_i$ breaks up into a finite set of isolated critical points,
and for each Betti number $b_j$ of $\CS_i$ there are $b_j$ critical points of index $j + \mu_i$.

This argument has a nice physical interpretation.
In many instances, it is hard to argue that the space of marginal deformations remains exactly marginal
once {\it e.g.} non-perturbative corrections are taken into account.
The relation \eqref{MorseBottineq} eliminates the need for such ``costly'' arguments,
at least as far as applications to RG flows and topology of $\CT$ are concerned.
Indeed, even if the space of marginal deformations $\CS_i$ is deformed while the strongest form of the $\CC$-theorem still holds,
it breaks up into a set of isolated CFTs which have exactly the same spectrum of $\mu (T_*)$ as the contribution
of the Poincar\'e polynomial $P(\CS_i)$ to \eqref{MorseBottineq}.
In other words, for each Betti number $b_j$ of $\CS_i$ there are $b_j$ conformal theories of index $\mu = j + \mu (\CS_i)$.

Sometimes, a weaker version of this argument can be useful: it does not provide as detailed information as \eqref{MorseBottineq},
but allows to generalize \eqref{muthma} to situations with marginal operators that are not necessarily exactly marginal.
In such cases, one simply needs to modify the definition \eqref{mudef} at the UV fixed point by counting not only relevant
operators but also marginal ones, {\it i.e.} replace the strict inequality $\Delta (\CO) < d$ by $\Delta (\CO) \le d$:
\be
\mu (T_{\text{UV}}) \; = \; \# (\CO~\text{with}~\Delta(\CO) \le d)
\label{murelmarg}
\ee
This leads to a variant of \eqref{muthma} that can be further refined if one knows which operators at
the fixed point $T_{\text{UV}}$ are marginally relevant and which are marginally irrelevant.

Again, the beauty of the relations \eqref{TviaRS}-\eqref{MorseBottineqx}
is that they have no explicit reference to the $\CC$-function and, therefore, can be tested in a much broader context.
A counterexample to any of these relations can be interpreted as an obstruction to the strongest version of the $\CC$-theorem.


\subsection{Supersymmetric RG flows}
\label{sec:SUSYflow}

Of particular interest are supersymmetric RG flows.
Until 20 years ago, very little was known about non-trivial superconformal fixed points in dimension $d > 2$ and RG flows between them. Then, following the pioneering work \cite{Seiberg:1994rs,Seiberg:1994pq} and related developments in string theory (most notably, the AdS/CFT correspondence \cite{Maldacena:1997re}) a lot of new superconformal theories have been discovered in the course of the past 20 years. Thanks to these developments, now we have a good supply of SCFTs in diverse dimensions and RG flows between them that can be used to test the ideas in the present paper.

When supersymmetry is maintained throughout the RG flow, the relevant operators come in complete multiplets and, as a result,
the index $\mu (T_*)$ is either an even number (in theories with four real supercharges), or a multiple of 4 (in theories with eight real supercharges).
Therefore, by analogy with $\hat c$ that, depending on the amount of supersymmetry, is a suitable multiple of the central charge in 2d SCFT,
it is convenient to introduce
\be
\hat \mu \; := \; \frac{\mu}{2} \qquad (4~\text{real supercharges})
\label{mu2def}
\ee
and
\be
\hat \mu \; := \; \frac{\mu}{4} \qquad (8~\text{real supercharges})
\label{mu4def}
\ee
In particular, in 3d SCFT the standard conventions are such that these two cases correspond to $\CN=2$ and $\CN=4$, respectively.
Therefore, in three dimensions we could simply write $\mu = \CN \cdot \hat \mu$.

This multiplicity is a familiar feature of supersymmetric theories. It has many different manifestations,
including {\it e.g.} the fact that it is often convenient to use ``complex dimension'' or ``quaternionic dimension''
to count the dimension of moduli spaces or target spaces in supersymmetric theories.
This, in fact, is exactly the case in our setup since $\mu (T_*)$ is the dimension of a relevant manifold $\CR (T_*)$
which, depending on the amount of supersymmetry, can be a complex or quaternionic space.
Indeed, in such cases, the relevant deformations that parametrize $\CR (T_*)$ come in chiral multiplets or hypermultiplets,
so that we can write \eqref{TviaIandR} and \eqref{TviaRS} as
\be
\CT \; = \; \bigcup_{\CS_i~\text{conformal}} \CS_i \times \C^{\hat \mu_i} \qquad (\text{chiral multiplets})
\label{Tcomplex}
\ee
or
\be
\CT \; = \; \bigcup_{\CS_i~\text{conformal}} \CS_i \times \mathbb{H}^{\hat \mu_i} \qquad (\text{hypermultiplets})
\label{Thyper}
\ee
This cell-bundle decomposition of the theory space $\CT$ has an important implication for the Poincar\'e polynomial $P(\CT)$,
which in both of these cases simply equals the Morse-Bott polynomial, {\it cf.} \eqref{MorseBottineqx}:
\be
\boxed{ \phantom{\oint} P(\CT) \; = \; \sum_{\CS_i~\text{conformal}} q^{\mu (\CS_i)} \cdot P(\CS_i) \phantom{\oint} }
\label{PMBeq}
\ee
In other words, when all fixed points have even values of $\mu (T_*)$ (or multiple of 4, which is a special case),
then all coefficients $Q_k$ in \eqref{Morseineq} and \eqref{MorseBottineq} must vanish.

This is a special case of a basic principle in Morse theory --- called the {\it lacunary principle} of Morse ---
that any gap in the sequence $N_0$, $N_1$, $N_2, \ldots$ forces a relation between $MB(q)$ and $Q(q)$.
In particular, when $MB(q)$ contains only even powers of $q$ it follows from \eqref{MorseBottineqx} that
neither $P(q)$ nor $(1+q) Q(q)$ can contain odd powers of $q$ and we arrive at \eqref{PMBeq}.
As usual, we should point out that a counterexample to any of these conclusions is an obstruction to
the strongest version of the $\CC$-theorem, of course, assuming all the details are taken into account.

Another virtue of supersymmetric RG flows is that they provide additional tools for analyzing the behavior near the end-points of the flow,
which is precisely how we want to approach the study of moduli spaces, topology of $\CT$, {\it etc.}
In particular, one tool that can be useful for computing the index $\mu (T_*)$ at the critical point is the superconformal index $I(T_*)$.

In holographic RG flows, supersymmetry also leads to considerable simplifications and often yields
a system of first-order gradient flow equations (that follow from setting SUSY variations of fermions to zero).


\subsection{Symmetries}
\label{sec:equivariant}

As we mentioned earlier, it is often convenient to restrict our attention to a class of theories with certain symmetry properties.
The version of Morse theory developed by Bott is perfectly suited to deal with such situations and, in fact,
was largely motivated by applications to symmetric spaces, some of which led to spectacular results.

Let $G$ be a symmetry group acting on couplings $\lambda_i$ that parametrize space $\CT$ of theories that enjoy such
symmetry transformations.
In such cases, $\CC : \CT \to \R$ needs to be a $G$-invariant function and its critical points are
orbits of the form
\be
\CS_i \; = \; G / K_i
\label{SGorbits}
\ee
for some subgroups $K_i$,
which are the stabilizers of these critical orbits and, therefore, are global symmetries of the CFTs $T_* \in \CS_i$.
Note, the role of the quotient by the symmetry group $K_i$ is precisely the same as in \cite{Green:2010da,Kol:2002zt,Kol:2010ub},
and many familiar conformal manifolds of SCFTs with large amount of supersymmetry
are indeed symmetric coset spaces\footnote{In fact, such conformal manifolds are believed to contain infinitely many ``rational'' CFTs.}
of the form \eqref{SGorbits}.
We also point out that orbits $G / K$ are K\"ahler manifolds and their complex analogues, $G_{\C} / K_{\C}$, are hyper-K\"ahler
(see {\it e.g.} \cite{Gukov:2006jk} for a review of these and other facts about geometry and topology of coadjoint orbits).

When RG flow is a gradient flow of a $G$-invariant $\CC$-function, the relations described in the previous
subsections have a very powerful refinement, formulated in terms of the $G$-equivariant cohomology groups.
For example, a suitable generalization of \eqref{MorseBottineq} looks like
\be
\sum_{i,j} q^{\mu_i + j} \dim H_G^j (\CS_i ; \R) \; = \; (1+q) Q(q) + \sum_k q^k \dim H_G^k (\CT ; \R)
\label{MBequivariant}
\ee
where, as usual, $Q(q)$ is a polynomial (or, power series) with non-negative coefficients.
Now, even a simple non-degenerate critical point, $\CS = \text{point}$, makes a rather non-trivial
contribution to the equivariant Morse-Bott series \eqref{MBequivariant} since
\be
H^*_G (\text{point}) \; = \; H^* (BG)
\ee
where $BG$ is the classifying space of the group $G$.
For example, when $G=U(1)$ the classifying space $BG = \cp^{\infty}$ contributes to \eqref{MBequivariant}
the Poincar\'e series
\be
P (\cp^{\infty}) \; = \; \frac{1}{1-q^2}
\ee
More generally, for $G=U(N)$ we have
\be
P (BG) \; = \; \frac{1}{(1-q^2) \cdots (1 - q^{2N})}
\ee
Furthermore, the $G$-equivariant cohomology of any $G$-space has a rich algebraic structure of an $H^* (BG)$-module.

In the next section, we will encounter yet another, more interesting and seemingly unrelated appearance of classifying spaces.


\section{Counting RG walls}
\label{sec:RGwalls}

There are examples of RG flows where the RG scale $t$ can literally be interpreted as one of the space-time dimensions.
Although examples of such RG flows are relatively rare, they provide good intuition and a concrete realization
of the idea that RG flows can be viewed as domain walls interpolating between two vacua.
In such cases, one can actually interpret Figure~\ref{fig:RGflows2} as an illustration of a soliton (domain wall)
in a larger theory that interpolates between vacua $T_{\text{UV}}$ and $T_{\text{IR}}$.
In all such (admittedly, special) situations counting of RG flows and studying their moduli spaces \eqref{Mflows}
reduces to a familiar problem in soliton physics.

One class of such examples is based on the AdS/CFT correspondence \cite{Maldacena:1997re} and consists of holographic RG flows.
In these examples, fixed points of the RG flow are indeed realized as vacua of a gravitational theory in $d+1$ dimensions,
and the RG flow itself is a described by a gravitational soliton that interpolates between these vacua.
The holographic RG flows have been extensively studied in the literature and, with sufficient supersymmetry,
yield BPS solutions which are gradient flows, see {\it e.g.} \cite{Freedman:1999gp,Behrndt:2000pz}.
As we illustrate in section \ref{sec:2d}, they provide excellent arena for testing the relations \eqref{muthma}-\eqref{muthmc}
and \eqref{TviaRS}-\eqref{MorseBottineq}.

Another special class of examples where RG scale $t$ is one of the space-time directions involves the so-called RG walls,
which are similar to the holographic RG flows, except that the ``ambient theory'' is still $d$-dimensional and does not contain gravity.
(The direction $t$ in Figure~\ref{fig:RGflows2} should be interpreted as one of the $d$ space-time directions.)
Again, counting RG flows and studying their moduli spaces in these examples turns into a more familiar problem of counting RG walls.

The notion of RG walls was introduced and studied in \cite{Brunner:2007ur} in the context of 2d examples,
which will serve us well for illustrating the ``count'' of RG walls.
The basic idea of \cite{Brunner:2007ur} is to divide the $d$-dimensional space-time of a CFT $T_{\text{UV}}$
into several domains and turn on relevant perturbations that initiate the RG flow to $T_{\text{IR}}$ only in some domains.
Then, separating domains of $T_{\text{UV}}$ and $T_{\text{IR}}$ is the RG defect / domain wall,
for which the RG time $t$ in Figure~\ref{fig:RGflows2} is now one of the space-time dimensions.
Following \cite{Brunner:2007ur}, we study $\CN=2$ supersymmetric RG flows between minimal model orbifolds:
\be
A_{n-2} / \Z_n \qquad \leadsto \qquad A_{k-2} / \Z_k
\label{AASUSYmmflow}
\ee
with $k<n$ (that, for simplicity, we further assume to be relatively prime).
The generalization to counting of RG walls in 3d and 4d theories is fairly straightforward and will be discussed in the end of this section.

Much like $\CN=2$ minimal models mentioned in section~\ref{sec:topology}, the orbifold theories in \eqref{AASUSYmmflow}
admit a Landau-Ginzburg description with the superpotentials $W = Y^n$ and $W=X^k$, respectively.
In fact, the minimal model orbifold $A_{n-2} / \Z_{n}$ is precisely the mirror dual of the $A_{n-2}$ minimal model,
so that the RG flow \eqref{AASUSYmmflow} is related to \eqref{SUSYminLG} by mirror symmetry.
This can give us some intuition about what to expect in \eqref{AASUSYmmflow}.

On the mirror side \eqref{SUSYminLG}, the RG flows relate different Landau-Ginzburg theories with polynomial superpotential $W(\Phi)$,
so that the space of theories is basically the space of coefficients of $W(\Phi)$.
If, for a moment, we restrict our attention to degree-$n$ superpotentials,
then this space is $(n+1)$-dimensional, namely $\C^{n+1} \setminus \{ 0 \}$,
where we had to exclude the origin $\{ 0 \}$ since in $\CN=2$ minimal models and their deformations
the superpotential can not be trivial, $W(\Phi) \ne 0$.
Moreover, rescaling $W(\Phi)$ by a non-zero (complex) number does not affect the IR physics,
so that in our present example the space of couplings is $\cp^{n} = \big( \C^{n+1} \setminus \{ 0 \} \big) / \C^*$.
Taking the inductive limit as $n \to \infty$, we conclude that RG flows between $\CN=2$ minimal models belong to the space
\be
\CT = \cp^{\infty} \cong BU(1)
\ee
To be more precise, this is the space of F-term couplings.
As is customary in supersymmetric theories,
we view the entire space of couplings --- all of which are renormalized along the RG flow --- as a fibration over the space of superpotential couplings.
As long as the fiber of this fibration is contractible, one can project the space of all couplings to the base,
namely to the space of superpotential couplings without loosing any information,
and this will be our strategy in the present paper, also in applications to higher-dimensional theories.
This, in particular, answers the question posed in \cite{Vafa:1988ue}.

The identification of the space $\CT$ with the classifying space $BG$ --- in this case, for $G=U(1)$ --- is very intriguing
and we plan to explore it more fully in the future.
However, some aspects are already clear at the present stage.
For instance, the group $G=U(1)$ here is obviously the flavor
symmetry group of a free chiral multiplet (that acts on $\Phi$ by phase rotations).
Moreover, its complexification by D-terms gives precisely $G_{\C} = \C^*$ that we saw earlier, in \eqref{MflowsSUSYMM}
as the moduli space of `basic' RG flow between neighboring minimal models, $A_m$ and $A_{m-1}$.
Complexification of the global symmetry group also plays an important role in 4d $\CN=1$ flows \cite{Green:2010da,Kol:2002zt,Kol:2010ub},
essentially for the same reason: in both cases, it arises as an obstruction to conformal invariance.
We expect this to be the case for all other $G$: their classifying spaces $BG$ have rich and interesting topology
that encodes information about RG flows within the corresponding class of CFTs labeled by $G$ and a cohomology class of $\CT = BG$,
\be
m \in H^* (BG) \quad \Rightarrow \quad \text{CFT}_{G,m}
\ee

Another aspect of the RG flow \eqref{AASUSYmmflow} that we expect from its mirror \eqref{SUSYminLG} is that the moduli space
of RG walls that realize \eqref{AASUSYmmflow} is of complex dimension $(n-k)$, {\it cf.} \eqref{MflowsSUSYMM}.
In order to verify this and other properties, it is convenient to describe the RG flow \eqref{AASUSYmmflow}
by embedding it into a larger theory that contains both chiral fields $X$ and $Y$ used in the LG description of minimal model orbifolds:
\be
\text{2d } \CN=(2,2) \text{ GLSM:}
\qquad\qquad
\begin{array}{l@{\;}|@{\;}cc}
& X & Y \\\hline
U(1)_{\text{gauge}} & -n & k
\end{array}
\qquad\qquad
W = X^k Y^n
\label{MMlinsigmodel}
\ee
Without the superpotential $W$,
the geometry of the linear sigma-model is given by the solution to the D-term constraint \cite{Witten:1993yc,Vafa:2001ra}:
\be
-n |X|^2 + k |Y|^2 = t
\ee
modulo the $U(1)$ action
\be
(X,Y) \to (Xe^{-n\theta}, Ye^{k\theta})
\ee
Here, $t$ is the running FI coupling
\be
t = \zeta + (k-n) \log \Lambda
\ee
so that, for $k<n$, the UV fixed point ($\Lambda \to \infty$) is at $t = - \infty$
and the IR fixed point ($\Lambda \to 0$) is at $t = + \infty$.
This model was discussed in detail in the context of closed string tachyon condensation \cite{Vafa:2001ra}.

In the phase $t \ll 0 $, the chiral superfield $X$ has a large vev and $U(1)$ is spontaneously broken to $\Z_n$.
Taking into account the superpotential $W$ we recover the Landau-Ginzburg description of the minimal model orbifold $A_{n-2} / \Z_n$.
On the other hand, when $t \gg 0$, the field $Y$ has a vev and $U(1)$ is spontaneously broken to $\Z_k$,
so that we recover $A_{k-2} / \Z_k$.
Therefore, the RG flow between these two theories can be understood as the flow
in the space of complexified FI parameter from $t = - \infty$ to $t = + \infty$.

For simplicity, let is consider the `basic' RG flow \eqref{AASUSYmmflow} with $k=n-1$.
By embedding into the above linear sigma-model, it can be understood as a flow in the space of the parameter $t$,
which takes values in the cylinder, $\exp (t) \in \cp^1 \setminus \{ 0, \infty \} \cong S^1 \times \R^1$.
This is precisely the complexification of $G = U(1)$ in the above discussion.
We can identify this cylinder with the moduli space of RG flows / RG walls,
$\CM (A_{n-2} / \Z_n, A_{n-3} / \Z_{n-1}) \cong \C^*$, {\it cf.} \eqref{MflowsSUSYMM}.
Indeed, as explained in \cite{Witten:1993yc}, for $k \ne n$ the quantum theory \eqref{MMlinsigmodel}
has no singularity on the $t$-cylinder and $(n-k)$ extra massive IR vacua with $\sigma \ne 0$,
so that the total index $\Tr (-1)^F$ remains invariant under the RG flow.

{}From the viewpoint of the RG wall, these extra $(n-k)$ vacua are trapped on the wall.
In the Landau-Ginzburg description of the minimal model orbifolds, the RG wall \eqref{AASUSYmmflow}
is given \cite{Brunner:2007ur} by the (equivariant) matrix factorization of the superpotential $W(X,Y) = Y^n - X^k$:
\be
M: \quad M_0 \xrightarrow{\quad p_0 \quad} M_1 \xrightarrow{\quad p_1 \quad} M_0 \,,
\qquad p_0 p_1 = W \cdot \text{Id}
\label{MMMfact}
\ee
The spectrum of chiral operators $\CH^* (M)$ on such a wall (not only RG wall) can be understood as a special case of
local operators that change the one-dimensional wall/defect from $M$ to $M'$.
In other words, such local operators are supported at points on the 1d defect,
such that to the left of a point one has the defect $M$ and to the right of it the defect is $M'$.
We need only a special case of this general setup since local operators that live on a defect $M$
(and don't change its type) correspond to $M'=M$.

Equivalently, using the so-called `folding trick', the local operators on $M$ can be described as
defect-changing local operators that can be placed between $M \otimes \bar M$ and a trivial defect.
In other words, they can be identified with local operators on which the tensor product defect $M \otimes \bar M$ can end \cite{Brunner:2007qu}.
Here, the dual defect, also known as ``anti-defect'' (or, ``anti-wall'') $\bar M$
is defined as a matrix factorization of the superpotential $-W$,
and is obtained from \eqref{MMMfact} simply by reversing the sign of $p_1$:
\be
\bar M: \quad M_0 \xrightarrow{\quad p_0 \quad} M_1 \xrightarrow{\,\,\,\, -p_1 \,\,\,\,} M_0
\ee
Now, using the standard rules (see {\it e.g.} \cite{Brunner:2007qu,Carqueville:2009ev}),
we can explicitly describe the fusion product $M \otimes \bar M$
of the wall $M$ and anti-wall $\bar M$:
\arraycolsep 6pt
\be
M \otimes \bar M: \quad
\left[\begin{array}{c}M_0 \otimes M_0 \\ \bigoplus \\ M_1 \otimes M_1 \end{array}\right]_0
\xrightarrow{\quad d^0\quad}
\left[\begin{array}{c} M_1 \otimes M_0 \\ \bigoplus \\ M_0 \otimes M_1 \end{array}\right]_1
\xrightarrow{\quad d^1 \quad}
\left[\begin{array}{c} M_0 \otimes M_0 \\ \bigoplus \\ M_1 \otimes M_1 \end{array}\right]_0
\label{MMbar}
\ee
where
\be
d^0=\left[\!\begin{array}{cc} p_0 \otimes \text{id} & \text{id} \otimes p_1 \\
\text{id} \otimes p_0 & p_1 \otimes \text{id} \end{array}\!\right]
\quad,\quad
d^1={\left[\!\begin{array}{cc} p_1 \otimes \text{id} & - \text{id} \otimes p_1 \\
- \text{id} \otimes p_0 & p_0 \otimes \text{id} \end{array}\!\right]}
\ee
Note, in general, the fusion product of defect lines described by matrix factorizations of the superpotentials $W_i$
is a matrix factorization of the potential $\sum_i W_i$. In particular, in our present case, since $M$ is described
by a matrix factorization of $W(X,Y) = Y^n - X^k$ while $\bar M$ is described by a matrix factorization of $-W(X,Y)$,
it is easy to see that \eqref{MMbar} is a matrix factorization of the zero superpotential.
Therefore, the differential $d$ in \eqref{MMbar} is a true (nilpotent) differential of a chain complex,
so that the spectrum of chiral operators $\CH^* (M)$ is simply the cohomology of this complex:
\be
\CH^i (M) \; = \; H^i (d) \; = \;
\frac{\text{ker}~ (d^i : (M \otimes \bar M)_i \to (M \otimes \bar M)_{i+1})}{\text{im}~ (d^{i-1} : (M \otimes \bar M)_{i-1} \to M \otimes \bar M)_i)}
\ee
where $i=0,1$ mod 2.
In particular, the supersymmetric index $\Tr (-1)^F$ that counts states localized on RG wall in this case is
\be
\Tr_{\text{wall}} (-1)^F = \dim \CH^0 (M) - \dim \CH^1 (M)
\label{wallindexx}
\ee

Similarly, by treating RG walls as ordinary domain walls, one can use the standard methods from soliton physics
to introduce analogues of \eqref{wallindexx} in a variety of other contexts.
For example, when RG walls are BPS objects, one can use the CFIV index \cite{Cecotti:1992qh,Fendley:1991ve,Cecotti:1992rm}
or its appropriate version to count RG walls:
\be
\Tr_{\text{wall}} \; (-1)^F \; = \; \chi \big( \CM_{\text{wall}} \big) \; = \; \chi \big( \CR (T_{\text{UV}}) \cap \CI (T_{\text{IR}}) \big)
\label{wallindex}
\ee
In 3d and 4d theories, concrete examples of counting domain walls that interpolate between different sectors of SUSY theories
have been studied in \cite{Gadde:2013wq} and \cite{Ritz:2002fm}.
In particular, in 3d $\CN=2$ theories one can actually refine the index \eqref{wallindex}
and study the flavored (or, equivariant) elliptic genus that counts degrees of freedom localized
on half-BPS walls that preserve $\CN=(0,2)$ supersymmetry on its two-dimensional world-volume \cite{Gadde:2013wq}:
\be
\Tr_{\text{wall}} \; (-1)^F q^{H_L} x^{\text{flavor}}
\label{wallellgenus}
\ee


\section{$\mu$-theorem in 2d}
\label{sec:2d}

Since conformal theories and RG flows among them are best understood in two dimensions,
this is a natural starting point for testing the proposed relations \eqref{muthma}-\eqref{muthmc}.
In fact, in two dimensions Zamolodchikov \cite{Zamolodchikov:1986gt} proved that the trace anomaly, $c$,
is strictly decreasing and the RG flow is the gradient flow of the $c$-function, at least in conformal perturbation theory.
This suggests that Zamolodchikov's $c$-function should be interpreted as Morse-Bott function on the space of 2d theories $\CT$.
The Hessian of this function automatically obeys the strongest form of the $\mu$-theorem \eqref{muthmc}
due to the fact, used in \cite{Zamolodchikov:1986gt}, that in the vicinity of any 2d RG fixed point ({\it cf.} \eqref{CnearT} up to rescaling):
\be
c(T) \; = \; c(T_*) + 3 (\Delta_i - 2) (\lambda_i)^2 + \ldots
\ee
where, without loss of generality, we chose the critical point $T_*$ to be at $\lambda^i=0$.

There are many concrete examples of 2d conformal theories that can be used to test \eqref{muthma}-\eqref{muthmc}
and to explore the geography of the space of RG flows $\CT$ via relations in section~\ref{sec:topology}.
Among the simplest examples are two-dimensional (non-supersymmetric) minimal models $T_m$ labeled by $m=2,3,4, \ldots$.
These theories have central charge
\be
c (T_m) = 1 - \frac{6}{m(m+1)}
\ee
(The theory with $m=2$ is empty.)
Moreover, $T_m$ has $m-2$ relevant operators, so that
\be
\mu (T_m) = m-2
\ee
and from \eqref{muthmb} we expect $n$-dimensional space of flows between theories $T_m$ and $T_{m-n}$.

%

\subsection{Counting $\CN=(0,2)$ flows}

Supersymmetric cousins of these theories, namely $\CN=(2,2)$ minimal models
were already mentioned earlier in section~\ref{sec:topology}, {\it cf.} \eqref{SUSYminLG}-\eqref{MflowsSUSYMM}.
Indeed, supersymmetry often helps to maintain analytical control throughout the entire RG flow,
providing a good supply of interesting examples to test \eqref{muthma}-\eqref{muthmc}.
Half-way between non-supersymmetric ($\CN=0$) and $\CN=(2,2)$ theories are two-dimensional
models with $\CN=(0,2)$ supersymmetry. Such models are interesting for a number of reasons.
Their gauge dynamics, akin to that of 4d $\CN=1$ theories is very interesting and exhibits rich phase structure,
non-perturbative effects, {\it etc.}
Conformal theories that are of our prime interest here can be used as world-sheet theories for the heterotic string.
Yet, the RG flows between 2d $\CN=(0,2)$ SCFTs are largely unexplored
and our goal in this section is to start filling this gap,
in part motivated by the recent developments \cite{Gadde:2014ppa,Bertolini:2014ela}.

An intricate tree of RG flows can be obtained by considering IR fixed points of 2d $\CN=(0,2)$ SQCD
with gauge group $U(N_c)$ and the following matter content:
\begin{center}
\begin{tabular}{c|ccccc}
\multicolumn{6}{c}{~~2d~$\CN=(0,2)$~SQCD}  \tabularnewline[.2cm]
 & $\Phi$ & $\Psi$ & $P$ & $\Gamma$ & $\Omega$\tabularnewline
\hline
$U(N_{c})$ & $\square$ & $\overline{\square}$ & $\overline{\square}$ & $ $${\bf 1}$ & $\mbox{det}$\tabularnewline
$SU(N_{f})$ & $ $${\bf 1}$ & $\square$ & $ $${\bf 1}$ & $ $${\bf 1}$ & ${\bf 1}$\tabularnewline
$SU(N_{b})$ & $\overline{\square}$ & ${\bf 1}$ & ${\bf 1}$ & ${\square}$ & ${\bf 1}$\tabularnewline
$SU(2N_{c}+N_{f}-N_{b})$ & ${\bf 1}$ & ${\bf 1}$ & $\square$ & $\overline{\square}$ & ${\bf 1}$\tabularnewline
$SU(2)$ & ${\bf 1}$ & ${\bf 1}$ & ${\bf 1}$ & ${\bf 1}$ & $\square$\tabularnewline
\end{tabular}
\par\end{center}
where $\Phi$ and $P$ are $\CN=(0,2)$ chiral (bosonic) multiplets, whereas $\Psi$, $\Gamma$, and $\Omega$ are Fermi multiplets. In the ``conformal window'' $N_c \le N_b \le N_c + N_f$ the theory flows to an exactly solvable $\CN=(0,2)$ SCFT labeled by the triple $(N_c,N_f,N_b)$. For example, the exact superconformal R-charges of the ``$E_6$ theory'' with $(N_c,N_f,N_b) = (1,2,2)$ are
\be
R_\Phi = R_P = R_\Gamma = \frac{1}{3} \,,\qquad R_\Psi=R_\Omega=0
\ee
Turning on the relevant deformation $\CO = \Phi_2 \Psi_2$
initiates an RG flow to another theory in this class with $N_c = N_f = N_b = 1$ and
\be
R_P = R_\Gamma = \frac{1}{2} \,,\qquad R_{\Phi} = R_\Psi=R_\Omega=0
\ee
(In fact, the latter is a free theory of gauge-invariant ``mesons'' $\Psi' = \Phi_1 \Psi_1$.)
Since both theories are exactly solvable,
it is easy to verify directly that $\mu$ indeed decrees along the RG flow.

In the rest of this section we will focus on the most basic holographic RG flows between
2d conformal theories with $\CN=(0,2)$ supersymmetry.
These flows are described by the simplest gauged supergravity theory,
namely $\CN=2$ three-dimensional $U(1)$-gauged supergravity coupled to matter,
and can serve as a prototype for more interesting holographic RG flows in higher dimensions.



\subsection{Phases of 3d $\CN=2$ gauged supergravity}

The simplest way to obtain 3d $\CN=2$ supergravity coupled to matter is via dimensional reduction of
the analogous 4d system with $\CN=1$ supersymmetry, which is well known and is well described in many textbooks (see {\it e.g.} \cite{Wess:1992cp}).
Here, we choose to do something a little more interesting and consider much less known 3d $\CN=2$ gauged supergravity
coupled to matter that involves Chern-Simons coupling for the $U(1)$ gauge field and, therefore,
can not be obtained by reduction from four dimensions.
In fact, our discussion below easily extends to a larger class of matter-coupled supergravity theories with
arbitrary abelian groups, but we refrain from doing so here to avoid clutter.

The first examples of such theories were studied in \cite{Deger:1999st,Deger:2000as,AbouZeid:2001tu}
and then generalized in a systematic approach \cite{deWit:2003ja} (see \cite{deWit:2004yr} for a nice summary).
Below we explore the phase structure of such theories in a way
that parallels the study of abelian $\CN=2$ gauge theories in two dimensions \cite{Witten:1993yc}.
Indeed, the latter share a number of similar features with our theories here: the number of supersymmetries is the same,
the spaces of scalar fields and parameters are K\"ahler in both cases, {\it etc.}
There are, however, also some differences, the most important of which perhaps is that scalar potential in our
case is induced by gauging $U(1)$ isometry of the K\"ahler target space.

In three dimensions, $\CN=(p,q)$ extended supergravity is associated with the AdS$_3$ supergroup
\be
\text{OSp} (p|2) \times \text{OSp} (q|2)
\ee
that extends the isometry group $SO(2,2)$ of AdS$_3$.
(Note, the latter is not a simple group but rather the direct product of two $SL(2,\R)$ factors.)
Via AdS/CFT correspondence \cite{Maldacena:1997re}, maximally supersymmetric AdS$_3$ vacua
of such theories are dual to 2d conformal field theories with $\CN=(p,q)$ supersymmetry.
Here we will be interested in the case $p=0$ and $q=2$, whose dual CFT is 2d $\CN=(0,2)$ theory.
Such supergravity has two gravitini fields that we combine into a complex vector-spinor $\psi_{\mu}$.
The bosonic content includes the dreibein $e_{\mu}^a$ and the spin-connection field $\omega_{\mu}^{ab}$.

We wish to couple this supergravity theory to a sigma-model with a target space $M$ (not necessarily compact):
\be
\CL = \CL_{\text{SUGRA}} + \CL_{\text{matter}}
\label{L3dungauged}
\ee
The fields of the sigma-model are complex scalar fields $\phi^i$ and spinors $\chi^i$.
The scalar fields $\phi^i$ parametrize a target manifold $M$
endowed with a K\"ahler metric $g_{i \bar j} = \partial_i \partial_{\bar j} K$,
where $K (\phi, \bar \phi)$ is the K\"ahler potential.
The coupling to supergravity depends on the $SO(2)$ target space connection $Q_i (\phi, \bar \phi)$:
\be
Q_{i} (\phi, \bar \phi) = \frac{1}{2} \partial_{i} K(\phi, \bar \phi)
\quad, \quad
Q_{\bar i} (\phi, \bar \phi) = \frac{1}{2} \partial_{\bar i} K(\phi, \bar \phi)
\label{QQdef}
\ee
which enters, in particular, the supersymmetry transformations of the fermions.

The last and the final step in building the desired matter-coupled supergravity is
gauging a $U(1)$ isometry of the target manifold $M$.
In order to preserve $\CN=2$ supersymmetry,
the $U(1)$ isometry must be holomorphic, {\it i.e.} should leave the complex structure invariant.
In other words, the problem at hand naturally leads us to study a holomorphic $S^1$-action on
a K\"ahler manifold $M$ that preserves the K\"ahler structure.

Let $v$ denote the vector field on $M$ that generates this $S^1$-action.
We are interested in circle actions that have at least one fixed point on $M$.
Indeed, as we shall see below, fixed points of the circle action correspond to supersymmetric AdS$_3$
vacua of the matter-coupled supergravity theory and, via the standard holographic duality,
these are fixed points of RG flow in the dual 2d $\CN=(0,2)$ boundary theory.
Then, such circle action is automatically Hamiltonian, {\it i.e.} there exists a moment map $H : M \to \R$
such that
\be
\iota_v \omega = d H
\label{holomH}
\ee
where $\omega$ is the K\"ahler form on $M$.

The Killing vector field $v = v^{i} \partial_{i} + v^{\bar i} \partial_{\bar i}$ generates a K\"ahler transformation
\be
\delta K = v^i \partial_{i} K + v^{\bar i} \partial_{\bar i} K = f (\phi) + \bar f (\bar \phi)
\label{ffdef}
\ee
for some holomorphic function $f (\phi)$. A constant shift of this function, $f \to f + i \zeta$,
with a real-valued parameter $\zeta$ is interpreted as the Fayet-Iliopoulos term.
Such constant shifts act as $H \to H + \zeta$ on the moment map
\be
H = i v^{j} Q_{j} - i v^{\bar j} Q_{\bar j} - \frac{i}{2} (f - \bar f)
\label{mmapdef}
\ee

In order to gauge this $U(1)$ isometry, we introduce an abelian gauge field $A_{\mu}$ and
couple it to the system \eqref{L3dungauged} by extending the pure supergravity with a Chern-Simons term
and gauging a subgroup of the sigma-model isometries:
\be
\CL \quad \leadsto \quad \CL - \frac{g}{8} \epsilon^{\mu \nu \rho} A_{\mu} F_{\nu \rho}
+ g \CL_{\text{Yukawa}} + g^2 V
\label{3dgravL2}
\ee
Here, $g$ is the coupling constant and $V$ is the scalar potential on $M$ induced by gauging,
\be
V \; = \; 2 W^2 - g^{i \bar i} \partial_i W \partial_{\bar i} W
\label{3dgravVW}
\ee
which is expressed in terms of the real ``superpotential''
\be
W(\phi, \bar \phi) = (H + \zeta)^2 + W_0
\ee
A real constant $W_0$ corresponds to gauging the R-symmetry group $U(1)_R$.
After gauging, the supersymmetry variations are also modified by $g$-dependent terms.
In particular, the SUSY variations of the gravitini $\psi_{\mu}$ and fermions $\chi^i$ have the form:
\bea
\delta \psi_{\mu} & = & \big( {\mathcal D}_{\mu} - g W \gamma_{\mu} \big) \epsilon \label{3dflow} \\
\delta \chi^{i} & = & \big( {\mathcal D}_{\mu} \phi^{i} \gamma^{\mu} + g g^{i \bar i} \partial_{\bar i} W \big) \epsilon \nonumber
\eea
These variations vanish for field configurations that describe supersymmetric flows and vacua in 3d $\CN=2$ supergravity coupled to matter.
In particular, BPS ``walls'' that preserve 2-dimensional Poincar\'e symmetry
are described by the following general metric ansatz (in mostly plus metric conventions, {\it cf.} \cite{Deger:2002hv}):
\be
ds^2 = dr^2 + e^{2A(r)} \eta_{\mu \nu} dx^{\mu} dx^{\nu}
\label{holRG}
\ee
where $\eta_{\mu \nu}$ is the flat 2d Minkowski metric and we assume $\gamma^r \epsilon = \epsilon$.
With these conventions, the radial flow in 3d supergravity via AdS/CFT corresponds
to the RG flow in the 2d $\CN=(0,2)$ boundary theory with the renormalization scale $e^{A(r)}$.
In particular, up to a sign, the standardly used radial variable $r$ plays the role of the RG time $t = -r$,
so that $r \to + \infty$ in the UV and $r \to - \infty$ in the IR, see Figure~\ref{fig:RGflows2}.
All holographic RG flows described by this ansatz correspond to space-dependent profile of the scalar fields $\phi^i (r)$
and the metric \eqref{holRG} which is encoded in a single function $A(r)$.
Since the scalar fields parametrize the K\"ahler manifold $M$, we are led to identify
the target space of our sigma-model with the ``theory space'' $\CT$ of the boundary theory,
\be
\CT = M
\label{TMid}
\ee
which, in principle, may be infinite-dimensional.
Indeed, the scalar fields $\phi^i$ of 3d $\CN=2$ gauged, matter-coupled supergravity correspond
to spin-0 operators $\CO_i$ of the 2d $\CN=(0,2)$ boundary theory, so that one-dimensional motion
in the space of fields \eqref{TMid} corresponds to a one-dimensional motion in the space of 2d theories in this class.

Moreover, for a general holographic RG flow of the form \eqref{holRG},
the $\CC$-function of the $d$-dimensional boundary CFT$_d$ is \cite{Freedman:1999gp}:
\be
\CC = \frac{\CC_0}{(A')^{d-1}}
\label{holC}
\ee
In the present case, $d=2$ and the first equation in \eqref{3dflow} provides a relation between
the $\CC$-function (or, equivalently, $A'$) and the value of the superpotential $W$:
\be
A' - 2gW = 0
\label{AviaW}
\ee
The second equation in \eqref{3dflow} says that stationary points of the flow are critical points of $W$,
{\it i.e.} solutions to $\partial_ i W = 2 (H + \zeta) \partial_i H = 0$.
In particular, the supersymmetric anti-de Sitter space of radius $\ell$ corresponds to $A(r) = \frac{r}{\ell}$
and is a stationary point of the RG flow in the 2d $\CN=(0,2)$ boundary theory.
Since we are working with a two-derivative gravity theory with the Lagrangian \eqref{L3dungauged} and \eqref{3dgravL2},
the central charges $c_L$ and $c_R$ of the 2d $\CN=(0,2)$ SCFT are equal and given by the celebrated Brown-Henneaux formula:
\be
c_L = c_R = \frac{3 \ell}{2 G_N}
\ee

%
Now it is easy to verify that the holographic RG flows between different AdS$_3$ vacua in this $U(1)$ gauged $\CN=2$ supergravity
indeed satisfy \eqref{muthma}-\eqref{muthmc}.

\subsection{Toy model}

As we pointed out earlier, fixed points of the holomorphic $S^1$-action on the space $M$
(that we now identify \eqref{TMid} with the theory space $\CT$ of the boundary theory)
are automatically fixed points of the flow equations that follow from \eqref{3dflow}.
The behavior near an isolated fixed point is modeled on a basic example:
\be
\CT \; = \; \C^n
\label{toyflat}
\ee
with the standard K\"ahler form:
\be
\omega = \frac{i}{2} \sum_{k=1}^n dz_k \wedge d \bar z_k
\ee
and the K\"ahler potential
\be
K \; = \; \frac{1}{2} \sum_{k=1}^n |z_k|^2
\ee
Note, from \eqref{QQdef} we get $Q_i = \frac{1}{4} \bar z_i$ and $Q_{\bar i} = \frac{1}{4} z_i$.
A holomorphic $S^1$-action on $\CT = \C^n$ is
\be
(e^{i \theta}) \cdot (z_1, \ldots, z_n) = (e^{i w_1} z_1, \ldots, e^{i w_n} z_n)
\ee
where $w_k \in \Z \setminus \{ 0 \}$ are the weights ($1 \le k \le n$).
This action is generated by the vector field
\be
v = i \sum_{k=1}^n w_k (z_k \frac{\partial}{\partial z_k} - \bar z_k \frac{\partial}{\partial \bar z_k})
\ee
with components $v^k = i w_k z_k$ and $\bar v^k = -i w_k \bar z_k$.
Therefore, using \eqref{ffdef} and \eqref{mmapdef}
we find the moment map:
\be
H = - \frac{1}{2} \sum_{k=1}^n w_k |z_k|^2
\ee
and the real superpotential
\be
W = \left( \zeta - \frac{1}{2} \sum_{k=1}^n w_k |z_k|^2 \right)^2 + W_0
\ee
Note, that in a vacuum ({\it i.e.} at the stationary point $z_k=0$ of the holographic RG flow) we have $W = \zeta^2 + W_0$
and the AdS$_3$ radius is given by \eqref{AviaW}:
\be
\ell^2 = (2 g W)^{-2} = \frac{1}{4 g^2 (\zeta^2 + W_0)^2}
\ee

Our next goal is to compute the mass matrix for the complex scalar fields $z_k = x_k + i y_k$.
Differentiating the scalar potential \eqref{3dgravVW}
and setting $z_k = 0$ we find the (normalized) eigenvalues of the mass matrix
\be
m_i^2 \ell^2
= \left(\frac{\zeta w_i}{\zeta^2 + W_0} + 1\right)^2 - 1
\label{masseigentoy}
\ee
Note, all mass eigenvalues come in pairs since $\CN=2$ supersymmetry requires scalar fields to be complex valued.
We also point out that mass eigenvalues \eqref{masseigentoy} automatically obey the
Breitenlohner-Freedman bound, which for scalar fields in AdS$_{d+1}$ reads $m^2 \ell^2 \ge - \frac{d^2}{4}$.

The standard dictionary of the AdS/CFT correspondence \cite{Gubser:1998bc,Witten:1998qj}
relates the mass spectrum of scalar fields in AdS$_{d+1}$
with conformal dimensions of the corresponding operators $\CO_i$ in the dual CFT$_d$:
\be
\text{scalars} : \qquad
m^2 \ell^2 = \Delta (\Delta - d)
\ee
In particular, scalar fields with $m^2 < 0$ (resp. $m^2 > 0$) correspond to relevant (resp. irrelevant) operators of the boundary CFT:
\bea
m^2 < 0 & \quad\Leftrightarrow\quad & \text{relevant} \nonumber\\
m^2 > 0 & \quad\Leftrightarrow\quad & \text{irrelevant} \nonumber
\eea
In our present problem, from \eqref{masseigentoy} we see that, for small values $\frac{\zeta w}{\zeta^2 + W_0} \ll 1$,
the mass spectrum of scalar fields in AdS$_3$ vacuum is completely determined by
the weights $\{ w_i \}$ of the holomorphic $S^1$-action at a given fixed point $T_* \in \CT$,
so that the sign of $m_i^2 \ell^2$ is the same as that of $w_i$.
Therefore, in this case, the index \eqref{mudef} that counts relevant operators in a CFT $T_* \in \CT$
is equal to the Morse index of the perfect Morse-Bott function \eqref{holomH}.

It is easy to generalize our toy model here to RG flows in more interesting spaces $\CT$,
including infinite-dimensional theory spaces.
According to \eqref{3dflow}, all such flows are gradient flows,
which in the vicinity of a fixed point are modeled on our basic example \eqref{toyflat}.
For example, the simplest K\"ahler manifold $\CT$ that has non-trivial topology
and admits a holomorphic $S^1$-action is $\cp^1 = \C \cup \{ \infty \}$.
It has the Fubini-Study metric with the K\"ahler potential
$K = \frac{1}{2} \log \left( 1 + |z^2| \right)$
and the K\"ahler form
\be
\omega = \frac{i}{2} \frac{dz \wedge d \bar z}{(1 + |z|^2)^2}
\ee
In this metric, there is an obvious holomorphic circle action on $\CT = \cp^1$ with the moment map,
\be
H = - \frac{1}{2} \frac{|z|^2}{|z|^2+1}
\ee
and two critical points (at $z=0$ and $z=\infty$) that correspond to the AdS$_3$ vacua
with $\ell^{-1} = 2g (\zeta^2 + W_0)$ and $\ell^{-1} = 2g ((\zeta - \tfrac{1}{2})^2 + W_0)$.
The two critical points have $\mu_{\text{UV}} = 2$ and $\mu_{\text{IR}}=0$, and admit
a 2-parameter family of gravitational solitons (domain walls) that interpolate them.
The moduli space of such gravitational solitons,
\be
\CM = \C^* \cong S^1 \times \R
\ee
is parametrized by the center-of-mass (in the $r$-direction) and the angular variable
that determines the direction of the flow on $\CT = \cp^1$.
This agrees with the general conjecture \eqref{muthmb}.
If we represent $\cp^1$ as a 2-sphere,
\be
x_1^2 + x_2^2 + x_3^2 = 1
\ee
and use the polar coordinates $x_1 = \sin \theta \cos \psi$, $x_2 = \sin \theta \sin \psi$, $x_3 = \cos \theta$,
then we obtain a description of this model presented in \cite{Deger:2002hv}, with a slight generalization.
Indeed, in these coordinates the symplectic form is $\omega = d \cos \theta d \psi$
and the moment map looks like $H = \cos \theta$, so that
\be
W = (\zeta + \cos \theta)^2 + W_0
\ee
The two critical points (AdS$_3$ vacua) are now at $\theta = 0$ and $\theta=\pi$.

It would be interesting to study in detail RG flows in theories with parameter spaces $\CT = \cp^n$, $\cp^{\infty}$, {\it etc.},
as well as flows among recently discovered 2d $\CN=(0,2)$ theories~\cite{Gadde:2014ppa,Bertolini:2014ela}.
In these, as well as higher-dimensional examples of holographic RG flows, the theory space $\CT$ is simply
the space of scalar fields of the bulk (super)gravity.
In particular, non-trivial topology of $\CT$ is a signal for (new) conformal theories.


\section{$\mu$-theorem in 3d}
\label{sec:3d}

A simple example of isolated 3d $\CN=2$ CFT is a three-dimensional version of the Landau-Ginzburg model \eqref{SUSYminLG},
namely a theory of a chiral superfield $\Phi$ with the superpotential $W = \Phi^3$.
The superpotential interaction drives the theory to a stable fixed point, at which $\Delta (\Phi) = \frac{2}{3}$.
Viewed as $T_{\text{UV}}$, this theory has only one non-trivial relevant perturbation, 
which drives the theory to a trivial IR fixed point.
Clearly, in this example $\mu (T_{\text{UV}}) = 2$ and $\mu (T_{\text{IR}}) = 0$,
so that both \eqref{muthma} and \eqref{muthmb} hold.

In order to verify \eqref{muthmc}, we need to compute the Hessian of the appropriate $\CC$-function.
In three dimensions, the candidate for the latter is the free energy of the Euclidean theory on a 3-sphere \cite{Jafferis:2010un,Jafferis:2011zi}
(or, equivalently \cite{Casini:2011kv}, disk entanglement entropy):
\be
\CC (T) = - \log |Z_{S^3} (T)|
\ee
Hence, the Hessian of this quantity at a fixed point $T_*$ is related to a two-point function of the relevant perturbations $\CO_i$ and $\CO_j$.
We leave it to an interested reader to compare this quantity with \eqref{muthmc}.

More interesting examples of RG flows can be found by starting with 3d $\CN=2$ supersymmetric QED, that is a theory
of a $U(1)$ vector multiplet coupled to chiral multiplets $Q$, $\tilde Q$ of charge $+1$ and $-1$, respectively.
Supersymmetric RG flows are induced by turning on superpotential for gauge-invariant operators,
which in this case include $Q \tilde Q$ and vortex-creation (monopole) operators
$V_{\pm}$ with charge $\pm 1$ under topological $U(1)$ symmetry of $\CN=2$ SQED.
For example, a perturbation by the quartic superpotential
\be
\Delta W = m Q \tilde Q Q \tilde Q
\label{SQEDWQQQQ}
\ee
drives the theory to $\CN=4$ SQED.
At this point, however, we should remember that $\CN=2$ SQED has an exactly marginal deformation
that parametrizes a one-complex-dimensional conformal manifold which is expected to be $\cp^1$
(see \cite{Strassler:1998iz} for a discussion of possible alternatives):
\be
\CS_{\text{CFT}} (\text{SQED}) \; = \; \cp^1 \; \cong \; \{ 0 \} \cup \C_{\lambda^{-1}}
\label{SQEDconfmfld}
\ee
In other words, we are precisely in the situation of section~\ref{sec:Bott}.
Here, $\{ 0 \}$ denotes the original IR fixed point
and $\C_{\lambda^{-1}} = \C^*_{\lambda} \cup \{ \lambda = \infty \}$ corresponds to its perturbation that in SQED is represented by turning on the sextic superpotential \cite{Strassler:1998iz}:
\be
W = \lambda (Q^3 \tilde Q^3 + V_+^3 + V_-^3)
\label{SQEDdeformW}
\ee
The best way to see this is via 3d mirror symmetry that relates $\CN=2$ SQED to the so-called XYZ model \cite{Aharony:1997bx,Kapustin:1999ha}.
The latter is a theory of three chiral multiplets $X$, $Y$, $Z$ with the superpotential $W = XYZ$
and with the following identification of gauge-invariant operators:
\bea
X & \sim & Q \tilde Q \nonumber \\
Y & \sim & V_+ \label{SQEDXYZdict} \\
Z & \sim & V_- \nonumber
\eea
Using this dictionary we see that deformation of $\CN=2$ SQED by the superpotential coupling \eqref{SQEDdeformW}
on the XYZ side corresponds to a general family of cubic superpotentials:
\be
W = \lambda_0 XYZ + \lambda (X^3 + Y^3 + Z^3)
\label{XYZdeformW}
\ee
Moreover, using 3d mirror symmetry and the dictionary \eqref{SQEDXYZdict},
it is easy to see that the perturbation \eqref{SQEDWQQQQ} in the XYZ model
corresponds to the mass deformation $\Delta W = m X^2$ for the chiral superfield $X$.

When $\lambda=0$ in \eqref{SQEDdeformW}-\eqref{XYZdeformW}, integrating out $X$ in the presence of $\Delta W$
leaves behind a marginally irrelevant superpotential $W \sim (YZ)^2$.
Therefore, for $\lambda=0$ the perturbation by $\Delta W = m X^2$ induces the RG flow from XYZ model
to a theory of two free chiral multiplets $(Y,Z)$, or equivalently, a theory of a free $\CN=4$ hypermultiplet.
The latter, in turn, is precisely a free theory of vortex solutions in $\CN=4$ SQED,
which is the endpoint of RG flow induced by \eqref{SQEDWQQQQ}.
Recall, that in \eqref{SQEDconfmfld} $\lambda=0$ corresponds to $T_* = \{ 0 \}$.

Now, let us consider RG flows from $\CN=2$ SQED (or dual XYZ model) with $\lambda \ne 0$.
They all flow to the same IR fixed point, which consists of two (decoupled) copies of the basic
$\CN=2$ theory with the cubic superpotential $W = \Phi^3$
that we discussed in the very beginning of this section.\footnote{We thank M.~Strassler for helpful discussions on this point.}
Indeed, at least for small values of $\lambda$, one can first integrate out $X$ as in the above discussion.
This pushes the theory toward IR free fixed point of $Y$ and $Z$.
Then, the cubic superpotential $W = \lambda (Y^3 + Z^3)$ takes over and diverts the RG flow to a new fixed point,
which consists of two copies of $\Phi^3$ theory.

To summarize, the perturbation \eqref{SQEDWQQQQ} drives $T_{\text{UV}} \in \CS_{\text{CFT}} = \{ 0 \} \cup \C_{\lambda^{-1}}$
to one of the two isolated fixed points:
\be
T_{\text{IR}} \; = \;
\begin{cases}
\CN=4~\text{SQED} \,,  & T_{\text{UV}} = \{ 0 \} \\
T_{Y^3} \otimes T_{Z^3} \,, & T_{\text{UV}} \in \C_{\lambda^{-1}}
\end{cases}
\label{endofSQED}
\ee
depending on whether $T_{\text{UV}}$ is either $\{ 0 \}$ or in $\C_{\lambda^{-1}}$, in the notations of \eqref{SQEDconfmfld}.
Moreover, there is a two-parameter family of flows from the first fixed point (viewed as a theory of free $\CN=2$ multiplets $Y$ and $Z$)
to the second fixed point, parametrized by coefficients of relevant perturbations $Y^3$ and $Z^3$.
Therefore, combining this with the results in the beginning of this section, we conclude that the two fixed points in \eqref{endofSQED}
have $\hat \mu \equiv \frac{\mu}{2} = 4$ and $\hat \mu=2$, respectively.

Therefore, from \eqref{Tcomplex} we conclude that the sequence of RG flows starting from $\CN=2$ SQED goes through
the ``patch'' of the theory space with the cell decomposition
\be
\CT \; = \; \{ \text{pt} \} \; \cup \; \C^2 \; \cup \; \C^4 \; \cup \; \cp^1 \times \C^5
\label{TSQED}
\ee
which, according to \eqref{TviaflowM}, should be viewed a convenient geometric way to package the information about
moduli spaces of RG flows.
Similarly, the Poincar\'e polynomial \eqref{PMBeq} of the space $\CT$ of RG flows,
\be
P (\CT) \; = \; 1 + q^4 + q^8 + q^{10} + q^{12}
\ee
is a convenient way to package the topology of conformal manifolds and values of $\mu (T_*)$.


\subsection{Index $\mu$ from the superconformal index}

As we mentioned earlier, one bonus feature of supersymmetry is that it allows to define
a refinement of the Witten index that takes into account the superconformal R-charge at the RG fixed point.
Thus, in 3d $\CN=2$ superconformal theory such index can be defined as \cite{Bhattacharya:2008zy}:
\be
I (T_*;q) \; = \; \Tr \; (-1)^F q^{\frac{R}{2} + j_3}
\label{3dindex}
\ee
possibly with additional fugacities for flavor symmetries, similar to \eqref{wallellgenus}.
This index receives contribution only from the states with $\Delta = R + j_3$,
so that spinless chiral operators of conformal dimension $\Delta$ contribute to the $q^{\frac{\Delta}{2}}$ term in \eqref{3dindex}.
In particular, marginal operators contribute to the coefficient of $q$,
while relevant operators contribute to the lower powers of $q$,
which by analogy with \cite{Dijkgraaf:2000fq,Gaberdiel:2008xb} one might call the ``polar part'' of $I(q)$.

For example, the superconformal index of a free chiral multiplet with R-charge $R_{\Phi}$ is \cite{Krattenthaler:2011da}:
\be
I (\Phi) \; = \; \frac{(q^{1-R_{\Phi}/2}; q)_{\infty}}{(q^{R_{\Phi}/2}; q)_{\infty}}
\label{3dchiralindex}
\ee
where $(x,q)_{\infty} = \prod_{i=0}^{\infty} (1-x q^i)$ is the Pochhammer symbol.
Among other things, it illustrates why free theories are actually subtle and delicate from the perspective of our approach to counting RG flows.
The counting of relevant and marginal operators, as well as the index \eqref{3dchiralindex}, all depend on the R-charge assignment.
In that sense, isolated CFTs --- such as minimal models --- are most convenient for counting RG flows and reconstructing
the topology of the space $\CT$ of RG flows, then come theories with conformal manifolds as in section~\ref{sec:Bott}, {\it etc.}

{}From the general expression \eqref{3dchiralindex} one can easily obtain the index of the $\Phi^3$ theory
considered in the very beginning of this section
\be
I (\Phi^3~\text{theory}) \; = \; \frac{(q^{2/3}; q)_{\infty}}{(q^{1/3}; q)_{\infty}}
\; = \; 1 + q^{1/3} + q^{4/3} - q^{2} + q^{7/3} + q^{8/3} + \ldots
\label{3dindexphi3}
\ee
and the index of 3d $\CN=2$ SQED which has a fairly rich space of RG flows \eqref{TSQED}:
\be
I (\text{SQED}) \; = \; \frac{(q^{2/3}; q)_{\infty}^3}{(q^{1/3}; q)_{\infty}^3}
\; = \; 1 + 3 q^{1/3} + 3 q^{2/3} + q + 3 q^{4/3} + 6 q^{5/3} + \ldots
\label{3dindexXYZ}
\ee
and where one can clearly recognize contribution of three chiral multiplets (of the dual XYZ model) with $R_{\Phi}=\tfrac{2}{3}$.
In these expressions, the relevant perturbation of the $\Phi^3$ theory contributes to the $q^{1/3}$ term in \eqref{3dindexphi3}
and the marginal operator of the $\CN=2$ SQED contributes to the linear term in \eqref{3dindexXYZ},
although in general, of course, there can be cancelations since \eqref{3dindex} is only an index and not the trace (Poincar\'e polynomial).
Yet, in simple examples like the ones considered here the index can be a useful tool for counting relevant and marginal deformations.
Indeed, naively interpreting \eqref{3dindexphi3} one might conclude that the fixed point in question
has no marginal operators and one relevant perturbation,
while \eqref{3dindexXYZ} suggests one marginal deformation and six relevant operators.


\section{$\mu$-theorem in 4d}
\label{sec:4d}

\subsection{Counting $\CN=1$ flows}

The advantage of supersymmetry is that it relates the R-symmetry and the dilatation operator \eqref{dilopDO}.
At a fixed point of $\CN=1$ RG flow, both generators are in the same superconformal multiplet and their eigenvalues are related via
\be
\Delta (\CO) \; = \; \frac{3}{2} R (\CO) \; = \; 1 + \frac{1}{2} \gamma (\CO)
\label{N1DeltaR}
\ee
where $R(\CO)$ is the R-charge of a chiral operator $\CO$, and $\gamma (\CO)$ is its anomalous dimension.
If $\CN=1$ supersymmetry is maintained throughout the RG flow, the relation \eqref{N1DeltaR} still holds, although its use is more limited.
Another advantage of supersymmetry is that at fixed points of an RG flow,
the spectrum of chiral operators (including relevant and marginal ones) can be studied with the help of the superconformal index.
(We will not use this powerful tool in the analysis here.)

Here, we will be interested in supersymmetric RG flows induced by chiral operators.
As explained in section \ref{sec:topology}, such RG flows map out a complex manifold $\CT$
and therefore we shall use \eqref{mu2def} and \eqref{Tcomplex}.
As with other examples considered in this paper, it is much easier to test \eqref{muthma} than \eqref{muthmb} or \eqref{muthmc}.
A simple example, considered in \cite{Leigh:1995ep}, is a 4d $\CN=1$ flow between
two versions of $SU(3)$ SQCD with different number of flavors and superpotential interactions.
Namely, the theory $T_{\text{UV}}$ has $N_f=9$ chiral multiplets $Q^i$ in the fundamental representation
and $N_f=9$ chiral multiplets $\tilde Q_i$ in the anti-fundamental representation, with the superpotential
\be
W_{\text{UV}} = \lambda_{\text{UV}} \left( Q^1 Q^2 Q^3 + Q^4 Q^5 Q^6 + Q^7 Q^8 Q^9 +
\tilde Q_1 \tilde Q_2 \tilde Q_3 + \tilde Q_4 \tilde Q_5 \tilde Q_6 + \tilde Q_7 \tilde Q_8 \tilde Q_9 \right)
\ee
This theory has a non-trivial conformal manifold $\CS_{\text{UV}}$ which at least contains a curve $\gamma_Q (g,\lambda_{\text{UV}}) = 0$.
(All chiral matter multiplets have the same anomalous dimension in this theory.)
Perturbation by the relevant operator $\CO = m (Q^3 \tilde Q_3 + Q^6 \tilde Q_6 + Q^9 \tilde Q_9)$
gives a mass to three flavors and initiates an RG flow to the IR theory $T_{\text{IR}}$,
which is $SU(3)$ SQCD with $N_f = 6$ and the quartic superpotential
\be
W_{\text{IR}} = \lambda_{\text{IR}} \left( Q^1 Q^2 \tilde Q_1 \tilde Q_2 + Q^4 Q^5 \tilde Q_4 \tilde Q_5 + Q^7 Q^8 \tilde Q_7 \tilde Q_8 \right)
\ee
This theory also has a non-trivial conformal manifold $\CS_{\text{IR}}$.
Hence, in order to examine the difference $\hat \mu_{\text{UV}} - \hat \mu_{\text{IR}}$, we need to use \eqref{murelmarg},
so that $\hat \mu_{\text{UV}}$ counts all chiral operators in $T_{\text{UV}}$ with $R (\CO) \le 2$
and $\hat \mu_{\text{IR}}$ counts all chiral operators in $T_{\text{IR}}$ with $R (\CO) < 2$.
Since the R-charge of the remaining six flavors is renormalized from $R = \frac{2}{3}$ to $R = \frac{1}{2}$ upon the RG flow,
the operators composed of these six flavors make equal contribution to $\hat \mu_{\text{UV}}$ and $\hat \mu_{\text{IR}}$.
Therefore, the difference $\hat \mu_{\text{UV}} - \hat \mu_{\text{IR}}$ precisely accounts for those relevant
operators in $T_{\text{UV}}$ that contain at least one of the three massive flavors (and are not in the spectrum of $T_{\text{IR}}$).\\

Now let us consider some interesting examples with dangerously irrelevant operators.
A simple, yet fairly rich class of $\CN=1$ RG flows was first studied in \cite{Kutasov:2003iy}
and later generalized and put in a broader context of Arnold's classification 
in \cite{Intriligator:2003mi}:
\be
\hat A \quad \leadsto \quad A_k \quad \leadsto \quad A_{k'}
\label{AAAflows}
\ee
Here, $k > k'$ and the theory $\hat A$ is the IR fixed point of
$\CN=1$ super-Yang-Mills theory with gauge group $SU(N_c)$ coupled to a chiral superfield $X$
in the adjoint representation and $N_f$ chiral multiplets in the fundamental representation, {\it cf.} Table~\ref{tab:adjSQCD}.
\begin{table}[htb]
\be
\begin{array}{l@{\;}|@{\;}ccc}
& Q_{i = 1, \ldots, N_f} & \tilde Q_{i=1, \ldots, N_f} & X \\\hline
SU(N_c) & \Box & \overline{\Box} & \text{adj} \\
SU(N_f) & \Box & {\bf 1} & {\bf 1} \\
SU(N_f) & {\bf 1} & \overline{\Box} & {\bf 1} \\
U(1)_R & y & y & \frac{1-y}{x}
\end{array}
\notag \ee
\caption{The field content of $\CN=1$ adjoint SQCD, the theory $\hat A$.}
\label{tab:adjSQCD}
\end{table}
\noindent
Then, each theory $A_k$ is obtained by adding to $\hat A$ a classical superpotential for the adjoint superfield $X$:
\be
A_k ~: \qquad W = \Tr X^{k+1}
\label{AkdefW}
\ee
Although such superpotential deformations (for $k>1$) are irrelevant in the UV adjoint SQCD,
upon the RG flow to the IR fixed point --- which is what we will be mostly interested in --- the R-charge of $X$ is renormalized
and some of these deformations become relevant, depending on the values of $N_c$, $N_f$, and $k$.
The chiral ring of theories $\hat A$ and $A_k$ is generated by the generalized ``mesons''
\be
Q^i X^n \tilde Q_j
\label{genmesons}
\ee
and single trace operators
\be
\Tr X^n
\label{xtozn}
\ee
To avoid unnecessary complications and extra relations among the chiral ring generators, we assume $N_c, N_f \gg 1$
while keeping the ratio $x := \frac{N_c}{N_f}$ fixed.
Since the $\CN=1$ adjoint SQCD is asymptotically free for $N_f < 2N_c$, we focus on the range of parameters:
\be
x = \frac{N_c}{N_f} > \frac{1}{2}
\label{xinadjSQCD}
\ee
Then, the first flow in \eqref{AAAflows} takes place for the following values of $x$:
\be
x_k < x \le k
\ee
where the upper bound comes from the stability of the vacuum
and the lower bound is the value of $x$ where \eqref{AkdefW} becomes a relevant deformation, $R(X^{k+1}) < 2$.
Explicitly,
\be
x_k \; = \;
\begin{cases}
\sqrt{\frac{1}{20}\left( \frac{(5k-4)^2}{9} + 1 \right)} \,,  & k<15\\
\frac{4 - \sqrt{3}}{6}k + \ldots \,, & k \gg 1
\end{cases}
\ee
Thus, $x_2 = \frac{1}{2}$, $x_3 = \frac{\sqrt{26}}{6} \simeq 0.850$, $x_4 = \frac{\sqrt{53}}{6} \simeq 1.213$, {\it etc.}



\begin{figure}[ht] \centering
\includegraphics[width=2.5in]{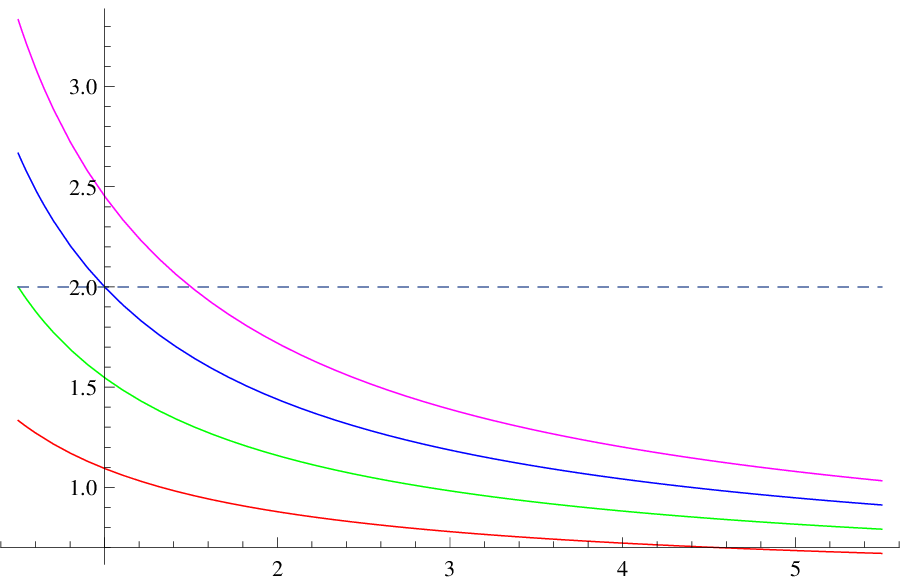}\qquad\qquad
\includegraphics[width=2.5in]{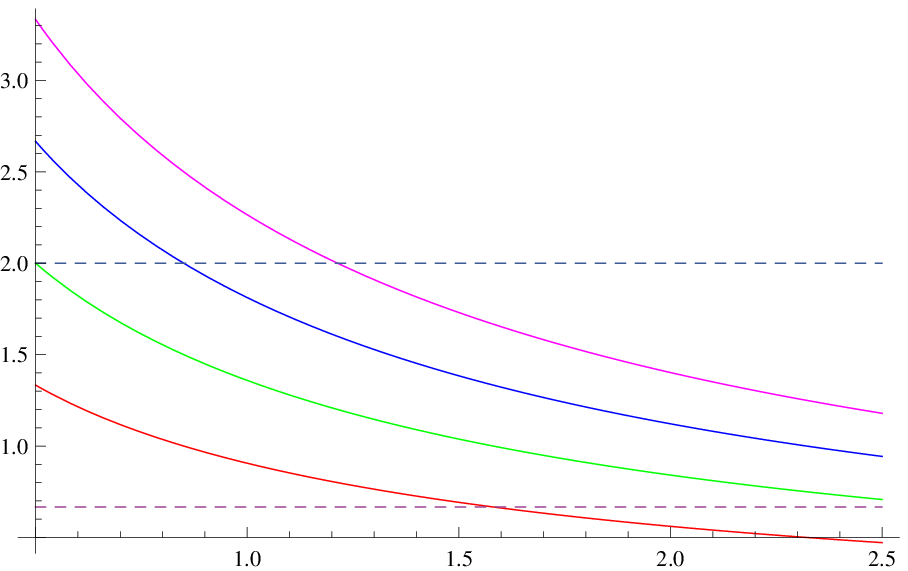}
\caption{\label{fig:SQCD}The spectrum of $\CN=1$ adjoint SQCD, the theory $\hat A$, as a function of $x = \tfrac{N_c}{N_f}$.
R-charges of generalized mesons $Q X^n \tilde Q$ for $n=0,1,2,3$ (Left) and
R-charges of $\Tr (X^n)$ for $n=2,3,4,5$ (Right). The dashed lines represent $R=2$ and $R=\tfrac{2}{3}$.}
\end{figure}

For larger values of $x \ge 3 + \sqrt{7}$, the generalized mesons \eqref{genmesons} of the $\hat A$ theory
hit the unitarity bound $R( \CO ) \ge \frac{2}{3}$ and the story gets more delicate (or, more interesting, depending on how one looks at it).
To keep things simple, we will limit our attention to the range $x \le 3 + \sqrt{7}$ where this does not happen
and $a$-maximization produces fairly simple expressions for the R-charges $R(Q)$ and $R(X) = \frac{1-R(Q)}{x}$ of the $\hat A$ theory \cite{Kutasov:2003iy}:
\be
y = R(Q) = R (\tilde Q) = \frac{3 + x(-3-6x+\sqrt{20x^2 - 1})}{3 - 6x^2}
\label{AhatyRQ}
\ee

Now let us take a closer look at the relevant and marginal operators in theories $\hat A$ and $A_k$
that determine $\mu (\hat A)$, $\mu (A_k)$, $\CS (\hat A)$, and $\CS (A_k)$,
which in turn determine topology of the space of RG flows.
At the $A_k$ fixed point:
\be
y = R(Q) = R (\tilde Q) = 1 - \frac{2x}{k+1}
\qquad , \qquad
R(X) = \frac{1-y}{x} = \frac{2}{k+1}
\label{RQQXAk}
\ee
so that
\be
R (Q X^n \tilde Q) = 2 + \frac{2n-4x}{k+1}
\qquad , \qquad
R(\Tr X^n) = \frac{2n}{k+1}
\ee
Note, the ``relevance'' of the generalized mesons does not depend on $k$ (only those with $n<2x$ are relevant),
but the dimension of their products does.
Therefore, at the second stage of the RG flow \eqref{AAAflows}, the theory $A_{k'}$ has the same list of
the generalized mesons as the parent theory $A_k$, but in general a different list of their products with $\Delta <3$.
Similarly, the list of relevant multi-trace operators made of $\Tr X^n$ depends on $k$ and, in general,
we have to consider products of \eqref{genmesons} and \eqref{xtozn}.

This brings us to an important point which apparently has not been previously discussed in the literature.
The theory $\hat A$ has no marginal deformations for generic values of $x$.
Indeed, operators $\CO$ which are products of \eqref{genmesons} and \eqref{xtozn} do not satisfy $R(\CO) = 2$
for generic values of $x$, especially if we recall that $x$ is a rational number \eqref{xinadjSQCD}
and R-charges of \eqref{genmesons}-\eqref{xtozn} take values in quadratic number fields.
However, a theory $A_k$ generically has lots of marginal deformations for {\it all} values of $x$.
The reason is that the R-charge \eqref{RQQXAk} of the adjoint superfield $X$ determined by
the superpotential \eqref{AkdefW} allows for marginal deformations by
\be
\delta W = \prod_i (\Tr X^{n_i}) \qquad \text{with} \qquad \sum_i n_i = k+1
\label{Akmarginal}
\ee
When used as perturbations of \eqref{AkdefW}, these have $n_i =2,\ldots,k-1$
since the gauge group is $SU(N_c)$ and $\Tr X^k$ is excluded by the chiral ring relation.
Consistent with these constraints, the first non-trivial example is a theory $A_3$, where one can consider a general quartic superpotential
\be
W \; = \; \lambda_0 \Tr X^4 + \lambda_1 (\Tr X^2)^2
\ee
Clearly, at least one of the coefficients $\lambda_j$ here needs to be non-zero in order to give \eqref{RQQXAk}.
We claim that \eqref{Akmarginal} are exactly marginal deformations and offer two proofs based on the techniques
of both \cite{Green:2010da} and \cite{Leigh:1995ep}.
Within the approach of \cite{Leigh:1995ep}, deformations \eqref{Akmarginal} are exactly marginal
when the anomalous dimension of the adjoint superfield $X$ satisfies $\gamma_X (g,\lambda_j) = \frac{6}{k+1}-2$.
Together with the $\beta$-function for the gauge coupling $g$, this equation cuts out a a codimension-1 hypersurface
in the space of couplings $\lambda_j$ (= coefficients of \eqref{Akmarginal}).
In the approach of \cite{Green:2010da} the proof is literally one line:
since there is no global continuous non-R symmetry acting on \eqref{Akmarginal} all such deformations are exactly marginal.
Therefore, we conclude that each theory $A_{k \ge 3}$
comes equipped with a conformal manifold $\CS (A_k)$
parametrized by the coefficients of exactly marginal deformations \eqref{Akmarginal}:
\be
\dim_{\C} \CS (A_3) = 1
\,, \qquad
\dim_{\C} \CS (A_4) = 1
\,, \qquad
\dim_{\C} \CS (A_5) = 3
\,, \qquad
etc.
\ee
For special values of $x$, additional combinations of the generalized mesons can yield extra marginal deformations
and increase the dimension of the conformal manifold $\CS (A_k)$.

As a concrete example, let us consider a cascading RG flow \eqref{AAAflows} with $k=4$ and $k' = 2$.
We take $x \gtrsim x_4$ just above the bound $x_k = \frac{\sqrt{53}}{6} \simeq 1.213$ for $k=4$.
Then, as is clear from Figure~\ref{fig:SQCD} and eq. \eqref{AhatyRQ}, the theory $\hat A$ has the following relevant operators:

$i)$ $Q X^n \tilde Q$ with $n=0,1,2$;

$ii)$ $\Tr X^n$ with $n=2,3,4,5$;

$iii)$ and the products  $(Q \tilde Q) (\Tr X^2)$, $(\Tr X^2) (\Tr X^2)$, $(\Tr X^2) (\Tr X^3)$.

\noindent
At the fixed point $A_4$, the R-charges are given by \eqref{RQQXAk},
and it is easy to verify that the list of relevant operators looks almost the same,
except for $\Tr X^4$ which becomes trivial in the chiral ring of the $A_4$ theory,
while $\Tr X^5$ and $(\Tr X^2) (\Tr X^3)$ become marginal.
So far \eqref{muthma} holds.

Upon turning on the relevant deformation $\CO = \Tr X^3$, one can continue the RG flow from $A_4$ to $A_2$.
The chiral ring of the latter theory is generated only by the generalized mesons \eqref{genmesons}
and at $x \simeq x_4$ has relevant operators $Q X^n \tilde Q$ with $n=0,1,2$.
The delicate feature of this stage of the RG flow is that mesons $M = Q \tilde Q$ hit the unitarity bound $R \ge \frac{2}{3}$.
It is believed that their R-charge is corrected to $R(M) = \frac{2}{3}$, so that $M$ decouple and become free fields.
If this is the case, the list of relevant operators in theory $A_2$ at $x \simeq x_4$
also includes $M^2$ and $(Q X \tilde Q)M$.
Curiously, this appears to violate \eqref{muthma} and, therefore, provides a concrete obstruction to
the strongest form of the $\CC$-theorem.
A careful reader may not find this too surprising since free theories are definitely most
subtle for applying the ideas of Morse-Bott theory.
Plus, if something discontinuous happens when $M$ become free fields, intuitively it is clear that
such phenomena are not likely to be captured by smooth gradient flows of the $\CC$-function.
However, as the next example shows, this is not the source of the problem.

Another illustrative example is the RG flow from the $\hat A$ theory to $\CN=1$ SQCD with no adjoints.
The IR limit of the latter theory is believed to be an interacting SCFT when $\frac{3}{2} N_c < N_f < 3N_c$
(or, equivalently, $\frac{1}{3} < x < \frac{2}{3}$) \cite{Seiberg:1994pq},
which we can call $A_1$ theory in the nomenclature \eqref{AkdefW} since the flow is induced by the relevant perturbation $\Tr X^2$.
Note, the range of $x \in (\tfrac{1}{3},\tfrac{2}{3})$ has a fairly modest overlap with \eqref{xinadjSQCD},
so without much loss of generality we can assume $x \gtrsim \frac{1}{2}$ just above the bound \eqref{xinadjSQCD}.
In this regime, the theory $\hat A$ is weakly coupled and the R-charges of $Q$, $\tilde Q$, and $X$ are just
a little under their canonical value of $R = \frac{2}{3}$.
This should not worry us since gauge-invariant operators \eqref{genmesons}-\eqref{xtozn} all obey the unitarity bound,
and the relevant ones are $Q \tilde Q$, $Q X \tilde Q$, $\Tr X^2$, and $\Tr X^3$.

Upon the flow from the $\hat A$ theory to $\CN=1$ SQCD with no adjoints (a.k.a. theory $A_1$),
the R-charges are renormalized to their new values, still given by \eqref{RQQXAk} with $k=1$:
\be
R(Q) = R (\tilde Q) = 1 - x
\ee
Note, for $x \in (\tfrac{1}{2},\tfrac{2}{3})$ the R-charges of all mesons $M = Q \tilde Q$ are above the unitarity bound
and the IR theory is an interacting SCFT.
However, the same phenomenon as we observed in a strongly coupled flow from $A_4$ to $A_2$ takes place:
meson mass operators, $\CO = M^2$, which were irrelevant in the UV become relevant in the IR.
Therefore, we find another interesting example of dangerously irrelevant operators and a topological
obstruction to the strongest form of the $\CC$-theorem.
Incidentally, precisely this type of behavior, where a double-trace operator crosses through marginality
is known \cite{Amariti:2012wc} to cause the ``branch flip'' in $a$-maximization
and is responsible for non-smooth behavior of the $a$-function under RG flow.
This strongly suggests that a violation of \eqref{muthma}-\eqref{muthmc} is indeed a signal for
violation of the strongest form of the $\CC$-theorem,
and that ``phase transitions'' along such $\CN=1$ RG flows are of second order.

To be completely fair, one needs to perform a more detailed study of moduli spaces in these theories
and the conformal manifolds $\CS (A_k)$ parametrized by \eqref{Akmarginal}.
Relegating further details to a separate publication, here we wish to point out two important features of this analysis.\footnote{We do not,
however, expect such analysis to change the conclusion about topological obstructions
to the strongest form of the $\CC$-theorem found here, and expect the gradient RG flows
to ``break'' precisely at the ``branch flip'' points.}
First, the entire RG flow \eqref{AAAflows} enjoys the symmetry group $G = SU(N_f) \times SU(N_f)$
whose action is summarized in Table~\ref{tab:adjSQCD}.
Therefore, we are in the setup of section~\ref{sec:equivariant} and need to use equivariant version of the Morse-Bott theory.
Moreover, the Higgs branch of these $\CN=1$ theories has a decomposition into ``Schubert cells'',
similar to the cotangent bundle of the Grassmannian $T^* Gr(N_c, N_f)$.
In fact, the latter is precisely the Higgs branch of the theory $\hat A$ with the cubic superpotential $W = Q X \tilde Q$.
Equivariant cohomology of the Grassmannian also plays an important role in a closely related theory (in three dimensions)
that categorifies gauged WZW model and the Verlinde formula.

It would also be interesting to extract the index $\mu (T_*)$ from
the superconformal index of $\CN=1$ theories $T_*$ along the lines of \cite{Beem:2012yn,Kutasov:2014wwa}.


\subsection{Counting $\CN=2$ flows}

Among 4d $\CN=2$ superconformal theories, the closest analogues of ``minimal models'' are the Argyres-Douglas theories \cite{Argyres:1995jj}
and their generalizations.
In these theories, computing $\mu (T_*)$ is relatively easy and boils down to counting lattice points inside the Newton
polygon of the local Seiberg-Witten geometry that describes the singularity type.
In other words, $\mu$ is essentially the Milnor number of the corresponding singularity.

Let us consider an example of the cascading RG flow passing through several Argyres-Douglas theories.
Namely, we consider the example studied in \cite{Xie:2013jc} keeping track of the value of $\mu (T_*)$ at each fixed point $T_*$:
\be
E_6 \quad \leadsto \quad D_4 \quad \leadsto \quad I_{3,2} \quad \leadsto \quad \text{Free hyper} \quad \leadsto \quad \text{Empty theory}
\label{Xieflows}
\ee
The $E_6$ theory, also known as $(A_1,E_6) \simeq I_{3,4} = (A_2,A_3)$, has
\be
E_6: \qquad x^3 + z^4 = 0\,, \qquad [x]=\frac{4}{7}\,, \qquad [z]=\frac{3}{7}\,, \qquad a=\frac{75}{56} \,, \qquad \hat \mu=6
\ee
where, following \cite{Xie:2013jc}, we record the local form of the Seiberg-Witten curve, dimensions of $x$ and $z$,
the anomaly coefficient $a$ and, most importantly, $\mu (T_*)$ using the shorthand notation $\hat \mu \equiv \frac{\mu}{4}$
introduced in \eqref{mu4def}.
The $E_6$ theory flows to the $D_4$ theory, a.k.a. $I_{3,3,} \simeq II_{3,3} = (A_1,D_4)$, which has
\be
D_4: \qquad x^3 + z^3 = 0\,, \qquad [x]=\frac{1}{2}\,, \qquad [z]=\frac{1}{2}\,, \qquad a=\frac{7}{12} \,, \qquad \hat \mu=4
\ee
The next critical point in the cascading RG flow \eqref{Xieflows} is the original Argyres-Douglas theory~\cite{Argyres:1995jj}:
\be
I_{3,2}: \qquad x^3 + z^2 = 0\,, \qquad [x]=\frac{2}{5}\,, \qquad [z]=\frac{3}{5}\,, \qquad a=\frac{43}{120} \,, \qquad \hat \mu=2
\ee
Flowing deeper in the IR, we arrive at a theory of a free hypermultiplet:
\be
\text{Free hyper:} \qquad x^3 + xz = 0\,, \qquad a=\frac{1}{24} \,, \qquad \hat \mu=1
\ee
The only relevant $\CN=2$ operator $\CO$ in this theory is the mass term for the hypermultiplet,
which drives the theory to the endpoint of the RG flow \eqref{Xieflows}:
\be
\text{Empty theory:} \qquad x^3 + z = 0\,, \qquad a=0 \,, \qquad \hat \mu=0
\ee
Clearly, \eqref{muthma} holds true at all stages of the flow,
and it would be interesting to directly verify \eqref{muthmb} and \eqref{muthmc}.

Using \eqref{Thyper} and \eqref{PMBeq}, we conclude that the cascading RG flow \eqref{Xieflows}
goes through the patch of the theory space with the following cell decomposition
\be
\CT \; = \; \{ \text{pt} \} \; \cup \; \mathbb{H} \; \cup \; \mathbb{H}^2 \; \cup \; \mathbb{H}^4 \; \cup \; \mathbb{H}^6
\ee
and the Poincar\'e polynomial
\be
P(\CT) \; = \; 1 + q^4 + q^8 + q^{16} + q^{24}
\ee
which provide a convenient geometric way to package the information
about moduli spaces of RG flows \eqref{TviaflowM} and the values of $\mu (T_*)$.


\acknowledgments{I would like to thank
N.~Deger, L.~Dixon, K.~Intriligator, V.~Lysov, Yu Nakayama, H.~Ooguri, L.~Rastelli, D.~Roggenkamp, E.~Sezgin, A.~Shapere,
M.~Strassler, R.~Sundrum, D.~Xie, and W.~Yan for useful discussions and comments.
It is also a pleasure to thank the organizers and participants of
the conference ``Progress and Application of Modern QFT'' in Aspen Feb. 16-21, 2015,
where I had the opportunity to discuss the results presented here.
This work is funded by the DOE Grant DE-SC0011632 and the Walter Burke Institute for Theoretical Physics.}


\newpage

\bibliographystyle{JHEP_TD}
\bibliography{RGflows}

\end{document}